\documentclass[aps,prd,twocolumn,showpacs,amsmath,amssymb]{revtex4-1}
\usepackage{amsmath}
\usepackage{mathtools}
\usepackage{graphicx}
\usepackage{subfigure}
\usepackage{epstopdf}
\usepackage{color}
\usepackage{multirow}
\usepackage{setspace}
\usepackage{overpic}
\usepackage{amssymb}
\usepackage[bookmarksnumbered, pdfstartview=FitH,colorlinks,urlcolor=blue, citecolor=blue,linkcolor=blue] {hyperref}
\usepackage{lineno}
\usepackage{bm}
\usepackage{rotating}
\usepackage[utf8]{inputenc}
\usepackage{morefloats}
\usepackage{hyperref}

\let\oldequation\equation
\let\oldendequation\endequation

\renewenvironment{equation}
  {\linenomathNonumbers\oldequation}
  {\oldendequation\endlinenomath}

\newcommand{\klnu}{D \to \bar K\ell^+\nu_{\ell}}
\newcommand{\kenu}{D^0\to K^-e^+\nu_e}
\newcommand{\kmunu}{D^0\to K^-\mu^+\nu_\mu}
\newcommand{\ksenu}{D^+\to \bar K^0e^+\nu_e}
\newcommand{\ksmunu}{D^+\to \bar K^0\mu^+\nu_\mu}
\newcommand{\koenu}{D^+\to \bar K^0e^+\nu_e}
\newcommand{\komunu}{D^+\to \bar K^0\mu^+\nu_\mu}
\newcommand{\ffK}{f^K_+(0)}

\begin{document}

\title{\boldmath Improved measurements of $D^0 \to K^-\ell^+\nu_\ell$ and $D^+ \to \bar K^0\ell^+\nu_\ell$}

\author{
M.~Ablikim$^{1}$, M.~N.~Achasov$^{4,c}$, P.~Adlarson$^{76}$, O.~Afedulidis$^{3}$, X.~C.~Ai$^{81}$, R.~Aliberti$^{35}$, A.~Amoroso$^{75A,75C}$, Q.~An$^{72,58,a}$, Y.~Bai$^{57}$, O.~Bakina$^{36}$
% ...
\\
\vspace{0.2cm}
(BESIII Collaboration)\\
\vspace{0.2cm} {\it
$^{1}$ Institute of High Energy Physics, Beijing 100049, People's Republic of China\\
% ...
$^{3}$ Bochum  Ruhr-University, D-44780 Bochum, Germany\\
$^{4}$ Budker Institute of Nuclear Physics SB RAS (BINP), Novosibirsk 630090, Russia\\
% ...
$^{35}$ Johannes Gutenberg University of Mainz, Johann-Joachim-Becher-Weg 45, D-55099 Mainz, Germany\\
$^{36}$ Joint Institute for Nuclear Research, 141980 Dubna, Moscow region, Russia\\
% ...
$^{57}$ Southeast University, Nanjing 211100, People's Republic of China\\
$^{58}$ State Key Laboratory of Particle Detection and Electronics, Beijing 100049, Hefei 230026, People's Republic of China\\
% ...
$^{72}$ University of Science and Technology of China, Hefei 230026, People's Republic of China\\
% ...
$^{75}$ University of Turin and INFN, (A)University of Turin, I-10125, Turin, Italy; (B)University of Eastern Piedmont, I-15121, Alessandria, Italy; (C)INFN, I-10125, Turin, Italy\\
$^{76}$ Uppsala University, Box 516, SE-75120 Uppsala, Sweden\\
% ...
$^{81}$ Zhengzhou University, Zhengzhou 450001, People's Republic of China\\
\vspace{0.2cm}
$^{a}$ Deceased\\
% ...
$^{c}$ Also at the Novosibirsk State University, Novosibirsk, 630090, Russia\\
% ...
}
}

\begin{abstract} Using 7.93 fb$^{-1}$ of $e^+e^-$ collision data
collected at the center-of-mass energy of 3.773 GeV with the BESIII
detector, we measure the absolute branching fractions of $\kenu$,
$\kmunu$, $\koenu$, and $\komunu$ to be $(3.521\pm0.009_{\rm
stat.}\pm0.016_{\rm syst.}) \%$, $(3.419\pm0.011_{\rm
stat.}\pm0.016_{\rm syst.}) \%$, $(8.864\pm0.039_{\rm
stat.}\pm0.082_{\rm syst.}) \%$, and $(8.665\pm0.046_{\rm
stat.}\pm0.084_{\rm syst.}) \%$, respectively. By performing a
simultaneous fit to the partial decay rates of these four decays, the
product of the hadronic form factor $\ffK$ and the modulus of the
$c\to s$ CKM matrix element $|V_{cs}|$ is determined to be
$\ffK|V_{cs}|=0.7171\pm0.0011_{\rm stat.}\pm0.0013_{\rm
syst.}$. Taking the value of $|V_{cs}|=0.97349\pm0.00016$ from the
standard model global fit or that of $\ffK=0.7452\pm0.0031$ from the
LQCD calculation as input, we derive the results
$\ffK=0.7366\pm0.0011_{\rm stat.}\pm0.0013_{\rm syst.}$ and
$|V_{cs}|=0.9623\pm0.0015_{\rm stat.}\pm0.0017_{\rm
syst.}\pm0.0040_{\rm LQCD}$.  \end{abstract}

\maketitle

\oddsidemargin  -0.2cm
\evensidemargin -0.2cm

\section{Introduction}

Improved measurements of semileptonic decays of charmed mesons provide
important inputs to further the understanding of weak and strong
interactions in the charm sector.  By analyzing their decay dynamics,
one can determine the product of the modulus of the
Cabibbo-Kobayashi-Maskawa (CKM) matrix element $|V_{cs(d)}|$ and the
hadronic transition form factor.  Taking $D\to \bar K e^+\nu_e$ as an
example, the hadronic transition form factors at zero-momentum
transfer
$f^{K}_+(0)$\cite{Lubicz:2017syv,Chakraborty:2021qav,Parrott:2022rgu,FermilabLattice:2022gku,Wu:2006rd,Verma:2011yw,Ivanov:2019nqd,Faustov:2019mqr,Ke:2023qzc}
can be calculated via several theoretical approaches, {\it e.g.},
lattice quantum
chromodynamics~(LQCD)~\cite{Lubicz:2017syv,Chakraborty:2021qav,Parrott:2022rgu,FermilabLattice:2022gku},
QCD light-cone sum rules (LCSR)~\cite{Wu:2006rd}, covariant
light-front quark model (LFQM)\cite{Verma:2011yw}, the covariant
confined quark model (CCQM) \cite{Ivanov:2019nqd}, and the
relativistic quark model (RQM) \cite{Faustov:2019mqr}.  Using the
value of $|V_{cs}|$ provided by the CKMFitter group~\cite{pdg2022},
the hadronic transition form factor $f_+^{K}(0)$ can be calculated,
resulting in a stringent test of the theoretical
predictions. Conversely, using the $f_+^{K}(0)$ value predicted by
theory allows the determination of $|V_{cs}|$, which is important to
test CKM matrix unitarity. Furthermore, measurements of the branching
fractions of $D^0 \to K^-\ell^+\nu_\ell$ and $D^+ \to \bar
K^0\ell^+\nu_\ell$~($\ell=e$ or $\mu$) are important to test lepton
flavor universality and isospin conservation in $D\to
K\ell^+\nu_\ell$.

Previously, the branching fractions of $D^0 \to K^-\ell^+\nu_\ell$ and
$D^+ \to \bar K^0\ell^+\nu_\ell$ were measured by
BESII~\cite{BES:2004rav,BES:2004obp,BES:2006kzp},
BaBar~\cite{BaBar:2007zgf}, Belle~\cite{Belle:2006idb},
CLEO-c~\cite{CLEO:2005rxg,CLEO:2005cuk,CLEO:2007ntr,CLEO:2009svp}, and
BESIII~\cite{BESIII:2021mfl,BESIII:2015tql,BESIII:2018ccy,BESIII:2017ylw,BESIII:2016hko,BESIII:2015jmz,BESIII:2016gbw}.
Studies of the decay dynamics of $\klnu$ were reported by BaBar~\cite{BaBar:2007zgf}, CLEO-c~\cite{CLEO:2009svp}, and BESIII~\cite{BESIII:2015tql,BESIII:2018ccy,BESIII:2017ylw,BESIII:2015jmz}.
The previous BESIII analysis used 2.93 fb$^{-1}$ of
$e^+e^-$ collision data taken at the center-of-mass energy $\sqrt
s=3.773$~GeV.  This paper reports improved measurements of the
branching fractions and decay dynamics of $D^0\to K^- \ell^+\nu_\ell$
and $D^+\to \bar K^0 \ell^+\nu_\ell$ by using 7.93~fb$^{-1}$ of
$e^+e^-$ collision data collected by the BESIII detector at $\sqrt
s=3.773$ GeV~\cite{Luminosity}.  Throughout this paper, charge conjugate modes are
implied.

\section{BESIII detector and Monte Carlo simulations}

The BESIII detector~\cite{Ablikim:2009aa} records symmetric $e^+e^-$ collisions
provided by the BEPCII storage ring~\cite{Yu:IPAC2016-TUYA01} in the center-of-mass energy range from 1.84 to 4.95~GeV, with a peak luminosity of $1.1 \times 10^{33}\;\text{cm}^{-2}\text{s}^{-1}$
achieved at $\sqrt{s} = 3.773~\text{GeV}$.
BESIII has collected large data samples in this energy region~\cite{Ablikim:2019hff,Li:2021iwf}. The cylindrical core of the BESIII detector covers 93\% of the full solid angle and consists of a helium-based
 multilayer drift chamber~(MDC), a time-of-flight system~(TOF), and a
 CsI(Tl) electromagnetic calorimeter~(EMC), which are all enclosed in
 a superconducting solenoidal magnet providing a 1.0~T magnetic
 field. The solenoid is supported by an octagonal flux-return yoke
 with resistive plate muon identification counters~(MUC) interleaved
 with steel. The main function of the MUC is to
 separate muons from charged pions, other hadrons and backgrounds
 based on their hit patterns in the instrumented flux-return yoke.
 The charged-particle momentum resolution at $1~{\rm
 GeV}/c$ is $0.5\%$, and the ${\rm d}E/{\rm d}x$ resolution is $6\%$
 for electrons from Bhabha scattering. The EMC measures photon
 energies with a resolution of $2.5\%$ ($5\%$) at $1$~GeV in the
 barrel (end-cap) region. The time resolution in the plastic
 scintillator TOF barrel region is 68~ps, while that in the end-cap
 region was 110~ps. The end-cap TOF system was upgraded in 2015 using
 multi-gap resistive plate chamber technology, providing a time
 resolution of 60~ps~\cite{etof}. Approximately 67\% of the data used
 here was collected after this upgrade.

Simulated samples produced with {\sc geant4}-based~\cite{geant4} Monte
Carlo (MC) software, which includes the geometric
description~\cite{Huang:2022wuo} of the BESIII detector and the
detector response, are used to determine detection efficiencies and to
estimate backgrounds. The simulation models the beam energy spread and
initial state radiation (ISR) in the $e^+e^-$ annihilations with the
generator {\sc kkmc}~\cite{ref:kkmc}.  Signal MC samples of the decays
$\klnu$ are simulated with a specific two-parameter series
expansion model~\cite{Becher:2005bg}.  The background is studied using
an inclusive MC sample that consists of the production of $D\bar D$
pairs from the $\psi(3770)$ (including quantum coherence for the
neutral $D$ channels), the non-$D\bar D$ decays of the $\psi(3770)$,
the ISR production of the charmonium states, and the continuum
processes. These processes are also generated with {\sc kkmc}.  The
known decay modes are modeled by {\sc evtgen}~\cite{ref:evtgen} with
branching fractions taken from the Particle Data
Group~(PDG)~\cite{pdg2022}, while the remaining unknown charmonium
decays are modeled with {\sc lundcharm}~\cite{ref:lundcharm}. Final
state radiation from charged final-state particles is incorporated
using {\sc photos}~\cite{photos}.

\section{Measurement Method}

At $\sqrt s=3.773$ GeV, the $D$ and $\bar D$ mesons are produced in
pairs from the $e^+e^-\to \psi(3770)\to D\bar D$ process, where $D$
stands for $D^0$ or $D^+$. This property allows us to do absolute
branching fraction measurement with the well established double-tag
(DT) method~\cite{DTmethod}.  In this method, the single-tag (ST)
candidate events are selected by reconstructing a $\bar D^0$ in the
six hadronic final states $\bar D^0 \to K^+\pi^-$, $K^+\pi^-\pi^0$,
$K^+\pi^-\pi^-\pi^+$, $K^+\pi^-\pi^0\pi^0$, $K^+\pi^-\pi^-\pi^+\pi^0$,
and $K^0_S\pi^+\pi^-$, or a $D^-$ in the six hadronic final states $D^-
\to K^{+}\pi^{-}\pi^{-}$, $K^0_{S}\pi^{-}$,
$K^{+}\pi^{-}\pi^{-}\pi^{0}$, $K^0_{S}\pi^{-}\pi^{0}$,
$K^0_{S}\pi^{+}\pi^{-}\pi^{-}$, and $K^{+}K^{-}\pi^{-}$.  These
inclusively selected candidates are referred to as ST $\bar D$ mesons.
In the presence of the ST $\bar D$ mesons, candidates for the signal
decays are selected to form DT events.  The branching fraction of the
signal decay is determined by \begin{equation} \label{eq:bf} {\mathcal
B}_{\rm sig}=N_{\mathrm{DT}}/(N_{\mathrm{ST}}^{\rm tot}\cdot
\bar\varepsilon_{\rm sig}), \end{equation} where $N_{\rm DT}$ is the
total DT yield in data, $N_{\mathrm{ST}}^{\rm tot}$ is the total ST yield  \begin{equation}
N_{\mathrm{ST}}^{\rm tot}=\sum_{i=1}^{6}N_{\mathrm{ST}}^{i},
\end{equation} where $N_{\rm ST}^i$ is the ST yield of tag mode
$i$, and $\bar{\varepsilon}_{\rm sig}$ is the weighted
efficiency of detecting the semileptonic decay, calculated by
\begin{equation} \bar\varepsilon_{\rm
sig}=\sum_{i=1}^{6}\frac{N_{\mathrm{ST}}^{i}\varepsilon_{\rm
sig}^i}{N_{\mathrm{ST}}^{\rm tot}}.  \end{equation}
Here, $\varepsilon_{\rm sig}^i=\varepsilon_{\rm DT}^i/\varepsilon_{\rm
ST}^i$, is the efficiency of detecting the semileptonic decay in the
presence of the ST $\bar D$ meson of tag mode $i$, where
$\varepsilon_{\rm DT}^i$ and $\varepsilon_{\rm ST}^i$ are the DT
efficiency and ST efficiency, respectively.

\section{\boldmath Selection of single tag $\bar D$ mesons}

For each charged track (except for those used for $K^0_S$
reconstruction), the polar angle ($\theta$) with respect to the MDC axis
is required to satisfy $|\cos\theta|<0.93$, and the point
of closest approach to the interaction point must be within 1\,cm in
the plane perpendicular to the MDC axis, $|V_{xy}|$, and within 10~cm
along the MDC axis, $|V_z|$.  Charged tracks are identified by using
the $\mathrm{d}E/\mathrm{d}x$ and TOF information, with which the
combined confidence levels under the pion and kaon hypotheses are
computed separately. Charged tracks are assigned to the particle type
that has the higher probability.

Candidate $K_S^0$ mesons are reconstructed from pairs of oppositely
charged tracks. For these two tracks, their polar angles are required
to satisfy $|\cos\theta|<0.93$ and the distance of closest approach to
the interaction point is required to be less than 20~cm along the MDC
axis. There is no requirement on the distance of closest approach in
the transverse plane, and no particle identification (PID) criteria
are required. The two charged tracks are constrained to originate from
the same vertex, which is required to be away from the interaction
point by a flight distance of at least twice the vertex
resolution. The quality of the vertex fit is ensured
by the requirement of $\chi^2 < 100$, and the invariant mass of the
$\pi^+\pi^-$ pair is required to be within $(0.487,0.511)$~GeV/$c^2$.

Neutral pion candidates are reconstructed via the
$\pi^0\to\gamma\gamma$ decay. Photon candidates are identified from
EMC showers. The EMC time difference from the event start time is
required to be within $(0,700)$~ns.  The energy deposited in the EMC
is required to be greater than 25~MeV in the barrel region
($|\cos\theta|<0.80$) and 50~MeV in the end-cap region
($0.86<|\cos\theta|<0.92$).  The opening angle between the photon
candidate and the nearest charged track in the EMC is required to be
greater than $10^{\circ}$. For any $\pi^0$ candidate, the invariant
mass of the photon pair is required to be within
$(0.115,0.150)$~GeV$/c^{2}$. To improve the momentum resolution, a
mass-constrained~(1C) fit to the known $\pi^{0}$ mass~\cite{pdg2022}
is imposed on the photon pair, and the $\chi^2$ of the 1C kinematic fit
is required to be less than 50. The four-momentum of the $\pi^0$
candidate from this kinematic fit is used for further analysis.

For the two-body tag mode of $\bar D^0\to K^+\pi^-$, the backgrounds
originating from cosmic rays, Bhabha and dimuon events are vetoed with
the following procedure defined in Ref.~\cite{deltakpi}.  It is
required that the TOF time difference between the two charged tracks
is less than 5~ns, and at least one EMC shower with energy greater
than 50~MeV or at least one additional charged track detected in the
MDC survives in each event. Further, it is required that the two
charged tracks are not consistent with being a muon pair or an
electron-positron pair, identified using the TOF,
$\mathrm{d}E/\mathrm{d}x$, EMC, and MUC measurement
information with the combined confidence levels $\mathcal{L}_{e}$,
$\mathcal{L}_{\mu}$, $\mathcal{L}_{K}$, and $\mathcal{L}_{\pi}$
for electron, muon, kaon, and pion hypotheses, respectively. To be
identified as an electron, $\mathcal{L}_{e}$ is required to be greater than 0 and
larger than $\mathcal{L}_{K}$, $\mathcal{L}_{\pi}$, as well as $0.8\cdot(\mathcal{L}_e+\mathcal{L}_\pi+\mathcal{L}_K)$.
To identify a track as a muon, $\mathcal{L}_{\mu}$ is required to be greater than 0, the deposited energy in the EMC
should fall within the range of 0.15 to 0.30 GeV, and the hit depth in the MUC needs to be either greater than
$(80\times|p_{\rm trk}|-60)$~{\rm cm} or greater than 40~{\rm cm}, where $p_{\rm trk}$ is the track momentum.

To identify the ST $\bar D$ mesons, we use two kinematic variables:
the energy difference $\Delta E\equiv E_{\bar D}-E_{\mathrm{beam}}$
and the beam-constrained mass $M_{\rm
  BC}\equiv\sqrt{E_{\mathrm{beam}}^{2}/c^{4}-|\vec{p}_{\bar
    D}|^{2}/c^{2}}$, where $E_{\mathrm{beam}}$ is the beam energy, and
$E_{\bar D}$ and $\vec{p}_{\bar D}$ are the total energy and momentum
of the ST $\bar D$ meson in the $e^+e^-$ center-of-mass frame,
respectively.  If there are multiple $\bar D$ candidates for a
specific tag mode, the one giving the least $|\Delta E|$ is chosen for
further analysis.  To suppress combinatorial backgrounds in the
$M_{\rm BC}$ distribution, tag dependent $\Delta E$ requirements are
imposed on the ST candidates.  The detailed $\Delta E$ requirements
and the ST efficiencies estimated by analyzing the inclusive MC sample
are summarized in Table~\ref{ST:realdata}.

\begin{figure*}[htbp]\centering
\includegraphics[width=0.98\linewidth]{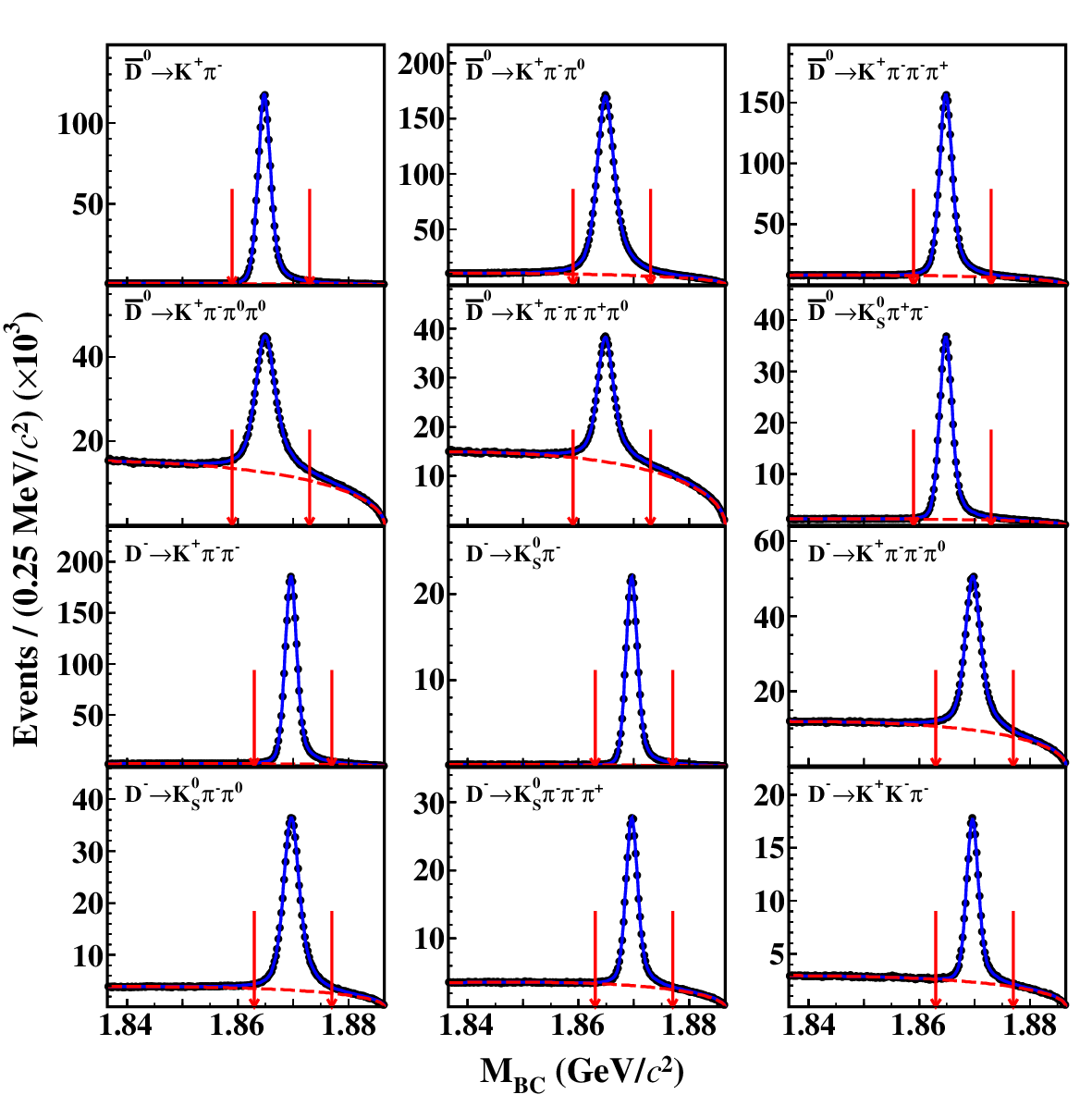}
\caption{ Fits to the $M_{\rm BC}$ distributions of the ST $\bar D$
  candidates.  The points with error bars are data, the blue curves
  are the best fits, and the red dashed curves are the fitted
  combinatorial background shapes. The pairs of red arrows show the
  $M_{\rm BC}$ signal window.
\label{fig:datafit_Massbc}}
\end{figure*}

For each tag mode, the yield of ST $\bar D$ mesons is obtained by
the maximum likelihood fit to the corresponding $M_{\rm BC}$ distribution.  In the fit, the
$\bar D$ signal shape is described by the sum of an MC-simulated signal shape, made by RooHistPdf in Rootclass~\cite{RootClass},
convolved with a double-Gaussian resolution function plus a single-Gaussian function, to account for
resolution difference between data and MC simulation and initial state radiation~(ISR) effects. The parameters of those functions are free. The background shape is
described by an ARGUS function~\cite{argus} with the endpoint fixed at
the $E_{\rm beam}$ value.  Figure~\ref{fig:datafit_Massbc} shows the
results of the fits to the $M_{\rm BC}$ distributions of the accepted
ST candidates in data for different tag modes. The candidates with
$M_{\rm BC}$ within $(1.859,1.873)$ GeV/$c^2$ for $\bar D^0$
tags and $(1.863,1.877)$ GeV/$c^2$ for $D^-$ tags are kept for
further analysis. Summing over the tag modes gives the total yields of
ST $\bar D^0$ and $D^-$ mesons ($N_{\rm ST}^{\rm tot}$) to be $(7922.7
\pm 3.4_{\rm stat.})\times 10^3$ and $(4135.4\pm2.4_{\rm stat.})\times
10^3$, respectively.

\begin{table}
\renewcommand{\arraystretch}{1.2}
\centering
\caption {The $\Delta E$ requirements, the ST $\bar D$ yields
  in data of the tag mode $i$ ($N_{\rm ST}^i$), and the ST
  efficiencies of tag mode $i$ ($\varepsilon_{\rm ST}^i$). The
  uncertainties are statistical only.  \label{ST:realdata}}
\resizebox{0.49\textwidth}{!}{
\begin{tabular}{c|c|c|c|c}
\hline
\hline
 $D$ & Tag mode & $\Delta E$~(MeV)  &  $N_{\rm ST}^i~(\times 10^3)$  &  $\varepsilon_{\rm ST}^i~(\%)    $       \\\hline
\multirow{6}{*}{$D^0$}
&$K^+\pi^-$                  &  $(-27,27)$ & $1452.5\pm1.3$&$65.28\pm0.01$\\
&$K^+\pi^-\pi^0$             &  $(-62,49)$ & $2908.1\pm2.0$&$35.69\pm0.01$\\
&$K^+\pi^-\pi^-\pi^+$        &  $(-26,24)$ & $1957.5\pm1.6$&$41.17\pm0.01$\\
&$K^+\pi^-\pi^0\pi^0$        &  $(-68,53)$ & $~698.8\pm1.3$&$15.24\pm0.01$\\
&$K^+\pi^-\pi^-\pi^+\pi^0$   &  $(-57,51)$ & $~457.1\pm1.1$&$16.49\pm0.01$\\
&$K^0_S\pi^+\pi^-$           &  $(-24,24)$ & $~448.7\pm0.7$&$37.60\pm0.01$\\
\hline
\multirow{6}{*}{$D^-$}
&$K^+\pi^-\pi^-$                  &  $(-25,24)$ & $2164.4\pm1.6$&$51.09\pm0.01$\\
&$K^{0}_{S}\pi^{-}$               &  $(-25,26)$ & $~249.4\pm0.5$&$50.72\pm0.02$\\
&$K^{+}\pi^{-}\pi^{-}\pi^{0}$     &  $(-57,46)$ & $~676.0\pm1.1$&$25.04\pm0.01$\\
&$K^{0}_{S}\pi^{-}\pi^{0}$        &  $(-62,49)$ & $~554.3\pm0.9$&$26.23\pm0.01$\\
&$K^{0}_{S}\pi^{-}\pi^{-}\pi^{+}$ &  $(-28,27)$ & $~304.0\pm0.7$&$29.53\pm0.01$\\
&$K^{+}K^{-}\pi^{-}$              &  $(-24,23)$ & $~187.3\pm0.5$&$41.12\pm0.02$\\
\hline
\hline
          \end{tabular}
          }

          \end{table}

\section{Selection of double tag events}

The candidates for $\kenu$, $\kmunu$, $\koenu$, and $\komunu$ are
selected from the remaining tracks in the presence of the ST $\bar D$
candidates.  Candidates for $K^-$ and $K_{S}^{0}$ are selected with
the same criteria as those used in the ST selection. The positron and
muon are identified using the TOF, $\mathrm{d}E/\mathrm{d}x$, and EMC
measurements with the combined confidence levels $\mathcal{L}_{e}$,
$\mathcal{L}_{\mu}$, $\mathcal{L}_{K}$, and $\mathcal{L}_{\pi}$, which
are calculated for electron, muon, kaon, and pion hypotheses,
respectively.  The positron candidate is required to satisfy
$\mathcal{L}_e > 0.8\cdot
(\mathcal{L}_e+\mathcal{L}_\pi+\mathcal{L}_K)$ and $\mathcal{L}_e
>0.001$.  The muon candidate is required to satisfy $\mathcal{L}_\mu >
\mathcal{L}_e$ and $\mathcal{L}_\mu >0.001$, and the energy of the muon
deposited in the EMC is required to be within $(0.1,0.3)$ GeV.

To suppress backgrounds associated with hadronic $D$ decays, it is
required that there are no additional good charged tracks on the
signal side ($N_{\rm extra}^{\rm trk}=0$).  To reject the backgrounds
from hadronic decays involving $\pi^0$, the maximum energy of extra
photons ($E_{\text{extra~}\gamma}^{\rm max}$) not used in the event
reconstruction is required to be less than 0.25 GeV.  Requirements on the
$\bar K$ and $\ell^+$~($M_{\bar K \ell}$) invariant mass, which are
$M_{K^-e^+}<1.83$~GeV/$c^2$ for $\kenu$, $M_{K^-\mu^+}<1.50$~GeV/$c^2$
for $\kmunu$, $M_{K_S^0e^+}<1.84$~GeV/$c^2$ for $\ksenu$, and
$M_{K_S^0\mu^+}<1.56$~GeV/$c^2$ for $\ksmunu$, are used to suppress
the backgrounds associated with the misidentification between $\pi^+$
and $\ell^+$.

The neutrino is not detectable by the BESIII detector.  In order to
determine the number of semileptonic $D$ candidates, we define
$U_{\mathrm{miss}}\equiv
E_{\mathrm{miss}}-|\vec{p}_{\mathrm{miss}}|c$, where
$E_{\mathrm{miss}}$ and $\vec{p}_{\mathrm{miss}}$ are the missing
energy and momentum of a DT event in the $e^+e^-$ center-of-mass
frame, respectively.  They are calculated by $E_{\mathrm{miss}}\equiv
E_{\mathrm{beam}}-E_{\bar K}-E_{\ell^+}$ and
$\vec{p}_{\mathrm{miss}}\equiv\vec{p}_{D}-\vec{p}_{\bar
  K}-\vec{p}_{\ell^+}$, where $E_{\bar K(\ell^+)}$ and $\vec{p}_{\bar
  K(\ell^+)}$ are the measured energy and momentum of the $\bar
K(\ell^+)$ candidate in an event. Here to improve the $U_{\mathrm{miss}}$ resolution,
$\vec{p}_{D}$ is evaluated as
$\vec{p}_{D} = -\hat{p}_{\bar D}
\sqrt{E_{\mathrm{beam}}^{2}/c^{2}-m_{\bar D}^{2} c^{2} }$, where
$\hat{p}_{\bar D}$ is the unit vector in the momentum direction of the
ST $\bar D$ meson and $m_{\bar D}$ is the known $\bar D$
mass~\cite{pdg2022}.

\section{Branching fractions}

\subsection{Results on branching fractions}
\label{bf_result}

\begin{figure*}[htbp]
\begin{center}
\subfigure{\includegraphics[width=0.8\linewidth]{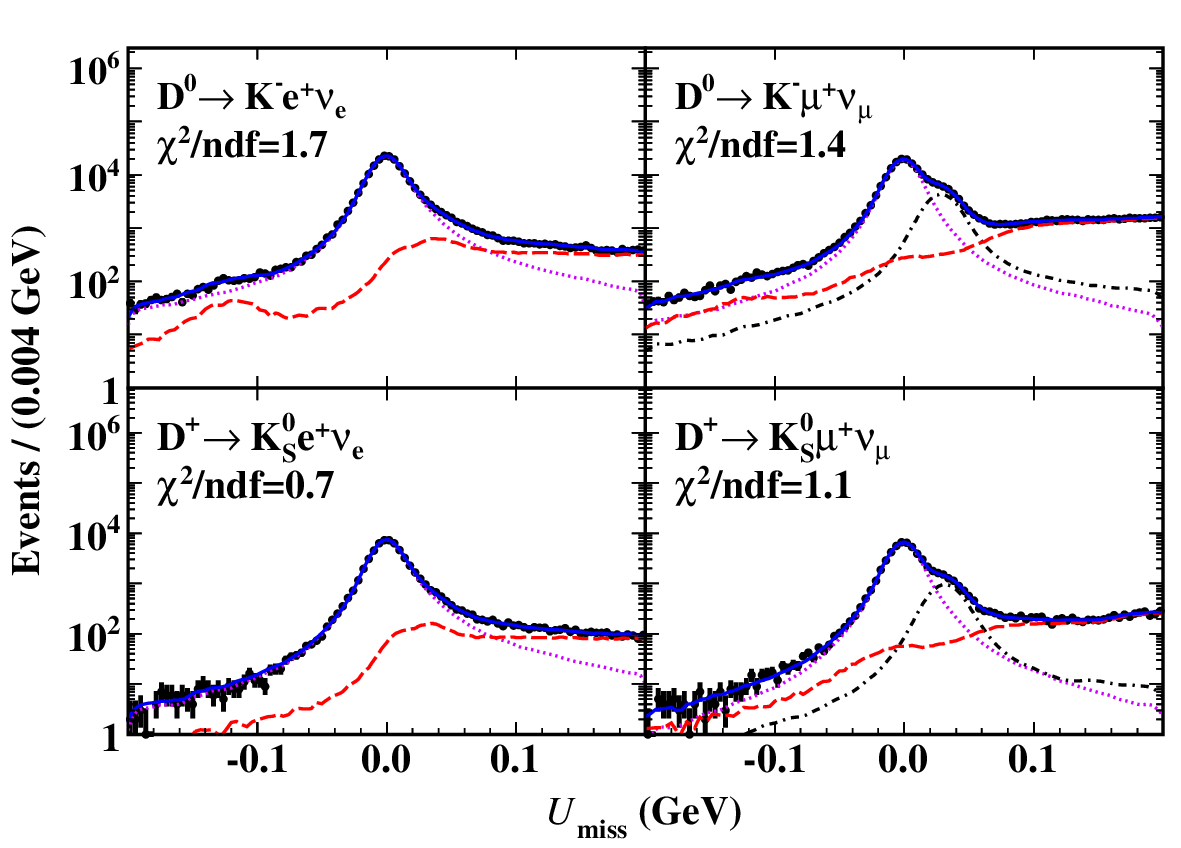}}

\caption{Fits to the $U_{\rm miss}$ distributions of the accepted candidates for $\klnu$ in data.
The points with error bars are data.
The violet dotted lines are the fitted signals.
The black dash-dotted lines are the fitted peaking backgrounds,
and the red dashed lines are the fitted combinatorial background shapes.
\label{fit_umiss}
}
\end{center}
\end{figure*}

After imposing all selection criteria, the $U_{\rm miss}$
distributions of the accepted candidates for $D\to \bar K\ell^+\nu_\ell$ in data are shown in Fig.~\ref{fit_umiss}.
Studies of the inclusive MC sample show that main backgrounds are caused mainly due to mis-identifying $\mu^+$ or $\pi^+$ as $e^+$ for $e$ channels; mis-identifying $\pi^+$ as $\mu^+$ for $\mu$ channels; or missing $\pi^0$(s) for all signal decays. 
The main sources of backgrounds, normalized yields and fractions, which are estimated from the inclusive MC sample, are listed in Table~\ref{umissbkg}.

For each signal decay, the signal yields in data are obtained from the maximum likelihood fits to the corresponding $U_{\rm miss}$ distributions. In the fit,
the signal shape is determined from the simulated shape convolved with a Gaussian function with free parameters, which accounts for different resolutions between data and MC simulation.  For the $D\to \bar K \mu^+\nu_\mu$ decays, the main peaking backgrounds $D\to \bar K \pi^+\pi^0$ are modeled by the simulated shapes convolved with the same Gaussian function as the corresponding signal, and their yields are free.
In the $U_{\rm miss}$ distributions,
there are also small peaking backgrounds from $D\to \bar K\mu^+\nu_\mu$ around 0.03~GeV for $D\to \bar Ke^+\nu_e$,
and small peaking background $D^0\to \pi^-e^+\nu_e$ around $-0.13$~GeV for $D^0\to K^-e^+\nu_e$.
Their shapes and yields are fixed in the fits and merged into other backgrounds in the plots.
The shapes of signal and all backgrounds are derived from signal and inclusive MC samples, respectively,
and all of them are made by RooHistPdf in Rootclass~\cite{RootClass}. From these fits, we obtain the signal yields of each signal decay in data~($N_{\rm DT}$).

The DT efficiencies are obtained by analyzing the corresponding signal MC samples. The obtained DT efficiencies and signal efficiencies for different signal decays in each tag mode as well as the weighted signal efficiencies ($\bar \varepsilon_{\rm sig}$) are listed in Table~\ref{tab:DT_efficiency}.

\begin{table}[htbp]
\caption{Main sources of backgrounds, normalized yields ($N_{\rm bkg}$) and fractions in all backgrounds~($f_{\rm bkg}$) for each signal decay estimated from the inclusive MC sample.}
\label{umissbkg}
\centering
\begin{tabular}{|c|l|c|c|}
\hline
\hline
Signal decay & Background source & $N_{\rm bkg}$ & $f_{\rm bkg}(\%)$ \\ \hline
\multirow{4}{*}{$\kenu$}
&$D^0 \to K^{*-}e^+\nu_e$      & 9820 & 46.7\\
&$D^0 \to K^{-}\mu^+\nu_\mu$   & 5421 & 25.8\\
&$D^0 \to K^{-}\pi^+\pi^0$     & 2756 & 13.1\\
&$D^0 \to \pi^{-}e^+\nu_e$     & 458 & 2.2\\
\hline
\multirow{3}{*}{$\kmunu$}
&$D^0 \to K^{-}\pi^+\pi^0$        & 44932 & 44.2\\
&$D^0 \to K^{-}\pi^+\pi^0\pi^0$   & 28815 & 28.3\\
&$D^0 \to K^{*-}\mu^+\nu_\mu$     & 8285  & 8.2\\
\hline
\multirow{3}{*}{$\ksenu$}
&$D^+ \to \bar K^{*0}e^+\nu_e$       & 2821 &  51.6\\
&$D^+ \to \bar K^{0}\mu^+\nu_\mu$    & 1558 &  28.5\\
&$D^+ \to K_S^0\pi^+\pi^0$           & 261  &  4.8\\
\hline
\multirow{3}{*}{$\ksmunu$}
&$D^+ \to K_S^0\pi^+\pi^0$            & 8760 &  50.3\\
&$D^+ \to \bar K^{*0}\mu^+\nu_\mu$    & 2546 &  14.6\\
&$D^+ \to  K_S^0\pi^+\pi^0\pi^0$      & 2072 &  11.9\\
\hline
\hline
\end{tabular}
\end{table}

\begin{table*}[htbp]
\centering\linespread{1.1}
\caption{The DT efficiencies $\varepsilon_{\rm DT}$, signal
  efficiencies $\varepsilon$ for different signal decays in each
  tag mode, as well as the weighted signal efficiencies $\bar
  \varepsilon_{{\rm sig}}$. The listed efficiencies are all in unit of \% and have taken into account corrections from data-MC differences originating from tracking and PID.
  For the $D^+$ signal decays, the efficiencies also include the branching
  fraction of $\bar K^0\to \pi^+\pi^-$. The uncertainties are
  statistical only.}  \small
\label{tab:DT_efficiency}
\resizebox{1.0\textwidth}{!}{
\begin{tabular}{c|cccc|c|ccccc }\hline \hline
\multicolumn{5}{c|}{$D^0$ decay} &         \multicolumn{5}{c}{$D^+$ decay} \\ \hline

Tag mode &$\varepsilon_{{\rm DT}, K^-e^+\nu_e}$&$\varepsilon_{K^-e^+\nu_e}$&$\varepsilon_{{\rm DT}, K^-\mu^+\nu_\mu}$&$\varepsilon_{K^-\mu^+\nu_\mu}$ &
Tag mode  &$\varepsilon_{{\rm DT}, \bar K^0e^+\nu_e}$&$\varepsilon_{\bar K^0e^+\nu_e}$&$\varepsilon_{{\rm DT}, \bar K^0\mu^+\nu_\mu}$&$\varepsilon_{\bar K^0\mu^+\nu_\mu}$ \\ \hline
$K^+\pi^-$                  &$43.45\pm0.03$&$66.57\pm0.05$&$34.76\pm0.03$&$53.26\pm0.04$&$K^+\pi^-\pi^-$                 &$8.14\pm0.01$&$15.94\pm0.02$&$6.78\pm0.01$&$13.26\pm0.02$\\

$K^+\pi^-\pi^0$             &$24.69\pm0.03$&$69.17\pm0.07$&$19.81\pm0.02$&$55.51\pm0.06$&$K^{0}_{S}\pi^{-}$              &$8.10\pm0.01$&$15.98\pm0.02$&$6.75\pm0.01$&$13.31\pm0.02$\\

$K^+\pi^-\pi^-\pi^+$        &$27.28\pm0.03$&$66.26\pm0.06$&$21.41\pm0.02$&$51.99\pm0.06$&$K^{+}\pi^{-}\pi^{-}\pi^{0}$    &$3.91\pm0.01$&$15.60\pm0.03$&$3.27\pm0.01$&$13.06\pm0.02$\\

$K^+\pi^-\pi^0\pi^0$        &$11.20\pm0.02$&$73.50\pm0.12$&$9.08\pm0.02$&$59.56\pm0.11$&$K^{0}_{S}\pi^{-}\pi^{0}$       &$4.11\pm0.01$&$15.67\pm0.02$&$3.48\pm0.01$&$13.28\pm0.02$\\

$K^+\pi^-\pi^-\pi^+\pi^0$   &$11.81\pm0.02$&$71.59\pm0.12$&$9.40\pm0.02$&$57.03\pm0.11$&$K^{0}_{S}\pi^{-}\pi^{-}\pi^{+}$ &$4.48\pm0.01$&$15.18\pm0.02$&$3.71\pm0.01$&$12.56\pm0.02$\\

$K^0_S\pi^+\pi^-$           &$24.89\pm0.03$&$66.19\pm0.07$&$19.59\pm0.02$&$52.07\pm0.06$&$K^{+}K^{-}\pi^{-}$             &$6.45\pm0.01$&$15.67\pm0.02$&$5.38\pm0.01$&$13.07\pm0.02$\\
\hline
$\bar \varepsilon_{{\rm sig}}$& &$68.32\pm0.03$& &$54.48\pm0.03$& $\bar \varepsilon_{{\rm sig}}$& &$15.78\pm0.01$& &$13.18\pm0.01$\\
\hline \hline
        \end{tabular}
        }
\end{table*}

With the signal yields in data~$N_{\rm DT}$, the weighted signal efficiencies~$\bar \varepsilon_{\rm sig}$, as well as the ST yield in data, the branching fractions of $\kenu$, $\kmunu$, $\ksenu$, and $\ksmunu$ are determined with Eq.~(\ref{eq:bf}) and listed in Table~\ref{bf_klnu}.

\begin{table*}[htp]
\centering
\caption{The signal yields in data $N_{\rm DT}$, the weighted signal
  efficiency $\bar \varepsilon_{\rm sig}$, as well as the branching
  fractions of the four signal decays $\mathcal B_{\rm sig}$.  For
  $\mathcal B_{\rm sig}$, the first uncertainties are statistical and
  the second are systematic.  For other quantities, the uncertainties
  are statistical only.  }
\label{bf_klnu}

\begin{tabular}{lcccc}
\hline\hline
Signal decay&$N_{\rm DT}$&$\bar \varepsilon_{{\rm sig}}~(\%)$&$\mathcal B_{\rm sig}~(\%)$\\ \hline
$\kenu$     &$190605\pm471$   & $68.32\pm0.03$   & $3.521\pm0.009\pm0.016$ \\
$\kmunu$    &$147596\pm488$   & $54.48\pm0.03$   & $3.419\pm0.011\pm0.016$ \\
$\ksenu$    &$~57846\pm256$   & $15.78\pm0.01$   & $8.864\pm0.039\pm0.082$ \\
$\ksmunu$   &$~47229\pm248$   & $13.18\pm0.01$   & $8.665\pm0.046\pm0.084$ \\ \hline\hline
\end{tabular}
\end{table*}

\subsection{Systematic uncertainties on branching fractions}
\label{section_b}

Table~\ref{table:bf_systot} summarizes the sources of the systematic uncertainties in the branching fraction measurements.
They are assigned relative to the measured branching fractions and are discussed below.

\paragraph{\boldmath \bf ST $\bar D$ yields}

The systematic uncertainty of the fits to the $M_{\rm BC}$ spectra is
estimated by varying the signal and background shapes and repeating
the fits for both data and the inclusive MC sample. A variation of the
signal shape is obtained by modifying the matching requirement between
generated and reconstructed angles from 15$^\circ$ to 10$^\circ$ or
20$^\circ$.  The uncertainty related to the background shape is
obtained by varying the endpoint by $\pm 0.2$ MeV.
In addition, the effect of removing the $M_{\rm BC}$ requirements from the ST selection is evaluated, and the difference with the nominal
measurement is taken as a systematic uncertainty accounting for possible mismodelling of the $M_{\rm BC}$ distribution in simulation.
Adding these three effects quadratically leads to a 0.3\% variation, which is taken as the
systematic uncertainty on $N_{\rm ST}$.  The uncertainty in the ST $\bar D^0$ yield is
correlated for $\kenu$ and $\kmunu$, while that for the ST
$D^-$ yield is correlated for $\ksenu$ and $\ksmunu$.

\paragraph{\boldmath \bf $K^-$ tracking and PID}

The $K^-$ tracking and PID efficiencies are studied by using a control
sample of hadronic $D\bar D$ events, with $D^0$ decaying into
$K^-\pi^+$, $K^-\pi^+\pi^+\pi^-$ and $\bar D^0$ decaying into $K^+\pi^-$,
$K^+\pi^-\pi^-\pi^+$, as well as $D^+$ decaying into $K^- \pi^+\pi^+$ and
$D^-$ decaying into $K^+\pi^-\pi^-$.  The ratios of the momentum weighted data and MC
efficiencies are $0.999\pm0.001$ and $1.000\pm0.001$ for tracking and
PID, respectively. The signal efficiencies are corrected by these
ratios, and their uncertainties, which are correlated for $\kenu$ and
$\kmunu$, are assigned as systematic uncertainties.

\paragraph{\boldmath \bf $K^0_S$ reconstruction}
\label{sec:gamma}

The $K^0_S$ reconstruction efficiencies are examined in two
aspects. The $\pi^\pm$ tracking efficiencies are determined by using
the control samples used in the studies of $K^-$ tracking and PID, but with a missing
$\pi^\pm$.  The efficiencies associated with the $K^0_S$ mass window
and $K^0_S$ decay vertex fit are examined using the
hadronic $D\bar D$ events, with $D^0$ or $D^+$ decaying into
$K^0_S\pi^+\pi^-$, $K^0_S\pi^+\pi^-\pi^0$, $K^0_S\pi^0$, $K^0_S\pi^+$,
$K^0_S\pi^+\pi^0$, and $K^0_S\pi^+\pi^+\pi^-$. The polar angle
distribution of the control sample is consistent with that in the
signal decays, therefore its effect on the $K^0_S$ reconstruction
efficiency is negligible. The momentum weighted difference between the
$K^0_S$ reconstruction efficiency of data and MC is 0.84\% for
$\ksenu$ and 0.88\% for $\ksmunu$, which are taken as the systematic
uncertainties.  These uncertainties are correlated for $\ksenu$ and
$\ksmunu$.

\paragraph{\boldmath \bf $\ell^+$ tracking and PID}
\label{sec:electron}

The tracking and PID efficiencies of $e^+$ and $\mu^+$ are studied by
using the control samples of $e^+e^-\to \gamma e^+e^-$ and $e^+e^-\to
\gamma \mu^+\mu^-$, respectively. The ratios of the data and MC efficiencies
weighted by momentum and cos$\theta$
are $0.999\pm0.001$ for $e^+$ tracking and $0.983\pm0.001$ for
$e^+$ PID; while they are $1.001\pm0.001$ for $\mu^+$ tracking and
$0.985\pm0.002$ for $\mu^+$ PID. The signal efficiencies are
corrected by these factors. After correction, the uncertainties of
ratios are assigned as the systematic uncertainties, and these
uncertainties are correlated for the four signal decays.

\paragraph{\bf MC model}

The detection efficiencies are estimated by using signal MC events
generated with the hadronic transition form factors measured in this
work.  The corresponding systematic uncertainties are estimated by
varying the parameters by $\pm1\sigma$.  These uncertainties are
independent for each signal decay.

\paragraph{\boldmath \bf $M_{\bar K \ell^+}$ requirement}
\label{m}

The uncertainty due to the $M_{\bar K\ell^+}$ upper bound in each
signal decay is studied by scanning the requirement from 1.74-1.84~GeV/$c^2$
for semielectronic decay and 1.46-1.56~GeV/$c^2$ for semimuonic decay
with a step of 0.01~GeV$/c^2$. We find the changes of
branching fractions $|\mathcal B_{\rm alternative}-\mathcal B_{\rm
  nominal}|$ are smaller than the statistical uncertainty difference
$\sqrt{|\sigma^2_{\rm alternative}-\sigma^2_{\rm
    nominal}|}$. Therefore, we neglect this systematic uncertainty.

\paragraph{\boldmath \bf $E_{\rm extra~\gamma}^{\rm max}$ and $N_{\rm extra}^{\rm trk}$ requirements}
\label{em}

The systematic uncertainty of the $E_{\rm extra~\gamma}^{\rm max}$ and
$N_{\rm extra}^{\rm trk}$ requirements is estimated by control samples of hadronic $D\bar D$ events,
with both $D$ and $\bar D$ decaying into one of the used ST hadronic final sates. The
difference of the acceptance efficiencies between data and MC
simulation is assigned as the systematic uncertainty.  These
uncertainties are correlated for $\kenu$ and $\kmunu$ or $\ksenu$ and
$\ksmunu$.

\paragraph{\boldmath \bf $U_{\rm miss}$ fit}

The systematic uncertainty due to the $U_{\rm miss}$ fit is considered
in two parts.  Since a Gaussian function is convolved with the
simulated signal shapes to account for the resolution difference
between data and MC simulation, the systematic uncertainty from the
signal shape is ignored.  The systematic uncertainty due to the
background shape is assigned by varying the relative fractions of
backgrounds from $e^+e^-\to q\bar q$ and the dominant background
channels in the inclusive MC sample within the uncertainties of their
input branching fractions.  The changes in the branching fractions are
taken as the corresponding systematic uncertainties.  These
uncertainties are independent for the four signal decays.

\paragraph{\bf MC statistics}

The relative uncertainties on the signal efficiencies are assigned as
systematic uncertainties due to MC statistics.  These uncertainties
are independent for the four signal decays.

\paragraph{\bf Quoted branching fractions}

For the $\ksenu$ and $\ksmunu$ decays, the uncertainty of the quoted
branching fraction of $K_{S}^{0}\to \pi^+\pi^-$ is
0.07\%~\cite{pdg2022}.  These uncertainties are correlated for
$\ksenu$ and $\ksmunu$.

\begin{table*}[htbp]
\caption{Relative systematic uncertainties (in \%) in the measurements of the branching fractions.
\label{table:bf_systot}}
\centering
\begin{tabular}{lcccc}
\hline
\hline

Source                              & $\kenu$ & $\kmunu$ & $\ksenu$ & $\ksmunu$\\\hline
$N_{\rm ST}$                       &0.30 &0.30 &0.30 &0.30   \\
$K^{-}$ tracking                   &0.10 &0.10 &--  &--   \\
$K^{-}$ PID                        &0.10 &0.10 &--  &--   \\
$K_S^0$ reconstruction              &--  &--  &0.85 &0.89 \\
$\ell^{+}$ tracking                &0.10 &0.10 &0.10 &0.10  \\
$\ell^{+}$ PID                     &0.10 &0.16 &0.10 &0.15  \\
MC model                            &0.20 &0.19 &0.10 &0.05   \\
$M_{\bar K \ell}$ requirement       &--   &--   &--   &--   \\
$E_{{\rm extra}~\gamma}^{\rm max}$ and $N_{\rm extra}^{\rm trk}$ requirement   &0.10 &0.10 &0.10 &0.10  \\
$U_{\rm miss}$ fit                  &0.18 &0.14 &0.06 &0.05  \\
MC statistics                       &0.05 &0.06 &0.07 &0.07  \\
Quoted branching fractions          &--   &--   &0.07 &0.07  \\

\hline

Total                               &0.46 &0.46 &0.93 &0.97  \\

\hline\hline

\end{tabular}
\end{table*}

\section{Hadronic transition form factors}

\subsection{Theoretical formula}

The differential decay width of the semileptonic decay $D \to \bar K \ell^+\nu_\ell$ can be expressed as~\cite{Faustov:2019mqr}
\begin{equation}
\label{ffk_function1}
\begin{array}{l}
   \displaystyle \frac{d\Gamma_{i}}{dq^{2}} =
   \frac{G_{F}^{2}|V_{cs}|^{2}}{24\pi^{3}}
   \frac{\left(q^{2}-m^{2}_{\ell}\right)^2|\vec p_{K}|}{q^{4}m^{2}_{D}}
   \displaystyle\left[\left(1+\frac{m^{2}_{\ell}}{2q^{2}}\right)m^{2}_{D}|\vec p_{K}|^{2}\right.\\\times|f^K_{+}\left(q^{2}\right)|^{2}
   \displaystyle\left.+\frac{3m^{2}_{\ell}}{8q^{2}}\left(m^{2}_{D}-m^{2}_{K}\right)^{2}|f^K_{0}\left(q^{2}\right)|^{2}\right],
\end{array}
\end{equation}
where $q$ is the four-momentum transfer to the $\ell^+\nu_\ell$ system, $|\vec p_{K}|$ is the modulus of the meson three-momentum in the $D$ rest frame, $G_F$ is the Fermi constant, $f^K_{+}(q^2)$ is the vector form factor, and $f^K_{0}(q^2)$ is the scaler form factor.
The series expansion~\cite{Becher:2005bg} is the most popular parameterization to describe
the hadronic transition form factor, which has the form
\begin{equation}
\begin{array}{l}
	\displaystyle f^{K}_{+}\left(q^2\right)=\frac{1}{P\left(q^2\right)\Phi\left(q^2\right)}\sum_{k=0}^{\infty}a_{k}
\left(t_0\right)\left[z\left(q^2,t_0\right)\right]^k.\\
	\end{array}
\end{equation}
Here, $a_{k}(t_0)$ are the real coefficients, and $P(q^2)=z(q^2,m_{D^{*+}_{s}}^{2})$, where $z(q^2,t_{0})=\frac{\sqrt{t_{+}-q^2}-\sqrt{t_{+}-t_{0}}}{\sqrt{t_{+}-q^2}+\sqrt{t_{+}-t_{0}}}$. The function $\Phi$ is given by
\begin{equation}
\begin{array}{l}
	\displaystyle \Phi(q^2)=\sqrt{\frac{1}{24\pi\chi_{V}}}\left(\frac{t_{+}-q^2}{t_{+}-t_{0}}\right)^{1/4}\left(\sqrt{t_{+}-q^2}+\sqrt{t_{+}}\right)^{-5}\\
	\displaystyle \times\left(\sqrt{t_{+}-q^2}+\sqrt{t_{+}-t_{0}}\right)\left(\sqrt{t_{+}-q^2}+\sqrt{t_{+}-t_{-}}\right)^{3/2}\\
	\displaystyle \times\left(t_{+}-q^2\right)^{3/4},
\end{array}
\end{equation}
where $t_{\pm}=(m_{D}\pm m_{K})^{2}$,
$t_{0}=t_{+}(1-\sqrt{1-t_{-}/t_{+}})$,
$m_{D}$ and $m_{K}$ are the masses of $D$ and $K$ particles,
$m_{D_s^{*+}}$ is the pole mass of the vector form factor $f^K_{+}(q^2)$ accounting for the strong interaction between
$D$ and $K$ mesons and usually taken as the mass
of the lowest lying $c\bar s$ vector meson $D_s^{*+}$, which is 2112.2~MeV~\cite{pdg2022}.
The $\chi_{V}$ parameter is obtained from dispersion
relations using perturbative QCD~\cite{chiV},
\begin{equation}
\begin{array}{c}
\displaystyle \chi_{V}=\frac{3}{32\pi^2m_c^2},
\end{array}
\end{equation}
where $m_c=1.27$~GeV is the $c$-quark mass.

In this analyses, the two-parameter series expansion is enough to describe the data, i.e.
\begin{equation}
\begin{array}{l}
	\displaystyle f^{K}_{+}\left(q^2\right)=\frac{1}{P\left(q^2\right)\Phi\left(q^2\right)}\left[a_{0}
\left(t_0\right)+a_{1}\left(t_0\right)z\left(q^2,t_0\right)\right].\\
	\end{array}
\end{equation}
By setting $r_1(t_0)=a_1(t_0)/a_0(t_0)$, the vector form factor $f^{K}_{+}(q^2)$ can be written as
\begin{equation}
\label{ffk_function2}
\begin{array}{l}
	\displaystyle f^{K}_{+}\left(q^2\right)=\frac{1}{P\left(q^2\right)\Phi\left(q^2\right)}\frac{f^{K}_{+}\left(0\right)P\left(0\right)\Phi\left(0\right)}{1+r_{1}
\left(t_{0}\right)z\left(0,t_{0}\right)}\\
	\displaystyle\times\left(1+r_{1}\left(t_{0}\right)\left[z\left(q^2,t_{0}\right)\right]\right).
	\end{array}
\end{equation}

The scalar form factor $f^K_{0}(q^2)$ is similar to $f^K_{+}(q^2)$ but with a one-parameter series expansion, which is given by~\cite{Faustov:2019mqr}
\begin{equation}
f^K_0(q^2) = \frac{1}{P(q^2)\Phi(q^2)}f^K_0(0)P(0)\Phi(0).
\label{equation:ff0}
\end{equation}
Here, $f^K_0(q^2)$ has the same normalization at $q^2=0$ as $f^K_+(q^2)$, {\it i.e.}
\begin{equation}
f^K_0(0) = f^K_+(0),
\end{equation}
but with a different pole mass $m_{D_{s0}^{*}(2317)^+}$ in $P(q^2)$, which is 2317.8~MeV~\cite{pdg2022}.

\subsection{Partial decay rates in data}

To obtain the hadronic transition form factors of the semileptonic
decays, the whole $q^2$ range is divided into 18 intervals for each
signal decay. The differential decay rate in the $i$-th $q^2$ interval
is determined as
\begin{equation}
	\frac{\mathrm{d}\Gamma_{i}}{\mathrm{d}q_{i}^{2}}=\frac{\Delta\Gamma_{i}}{\Delta q^{2}_{i}},
\end{equation}
where
	$\Delta\Gamma_{i}=N_{\mathrm{prd}}^{i}/(\tau_{D}\cdot N_{\mathrm{ST}}^{\rm tot})$
	is the partial decay rate in the $i$-th $q^2$ interval, $N_{\mathrm{prd}}^{i}$ is the number of events produced in the $i$-th $q^{2}$ interval,
and $\tau_{D}$ is the $D$ lifetime~\cite{pdg2022}.

In the $i$-th $q^{2}$ interval, the number of events produced in data is calculated as
\begin{equation}
	N_{\mathrm{prd}}^{i}=\sum_{j}^{N_{\mathrm{intervals}}}\left(\varepsilon^{-1}\right)_{ij}N_{\mathrm{DT}}^{j},
\end{equation}
	where $(\varepsilon^{-1})_{ij}$ is the element of the inverse efficiency matrix, obtained by analyzing the signal MC events.
The statistical uncertainty of $N_{\mathrm{prd}}^{i}$ is given by
\begin{equation}
\left[\sigma\left(N_{\mathrm{prd}}^{i}\right)\right]^2=\sum_{j}^{N_{\mathrm{intervals}}}\left(\varepsilon^{-1}\right)_{ij}^2\left[\sigma_{\rm stat}\left(N_{\mathrm{DT}}^{j}\right)\right]^2,
\end{equation}
where $\sigma_{\rm stat}(N_{\mathrm{DT}}^{j})$ is the statistical uncertainty of $N_{\mathrm{DT}}^{j}$.
The element $\varepsilon_{ij}^\alpha$ of the efficiency matrix with tag mode $\alpha$ is given by
\begin{equation}
	\varepsilon_{ij}^{\alpha}=\frac{N_{ij}^{\mathrm{rec}}}{N_{j}^{\mathrm{gen}}}\cdot \frac{1}{\varepsilon_{\mathrm{ST}}^{\alpha}},
\end{equation}
	where $N_{ij}^{\mathrm{rec}}$ is the number of events
        generated in the $j\text{-}$th $q^{2}$ interval and
        reconstructed in the $i$-th $q^{2}$ interval,
        $N_{j}^{\mathrm{gen}}$ is the number of events generated in
        the $j\text{-}$th $q^{2}$ interval, and
        $\varepsilon_{\mathrm{ST}}^\alpha$ is the ST efficiency with
        tag mode $\alpha$. The efficiency matrix elements
        $\varepsilon_{ij}$ weighted by the ST yields of data, which
        are presented in Tables 9$-$12 in \hyperref[appendix]{Appendix}, are given by
\begin{equation}
 \varepsilon_{\ij}=\sum_{\alpha=1}^{6}\frac{N_{\rm {ST}}^{\alpha}\varepsilon_{ij}^\alpha}{N_{\rm {ST}}^{\rm tot}}.
\end{equation}

\begin{figure*}[htbp]
\begin{center}
\subfigure{\includegraphics[width=0.98\textwidth]{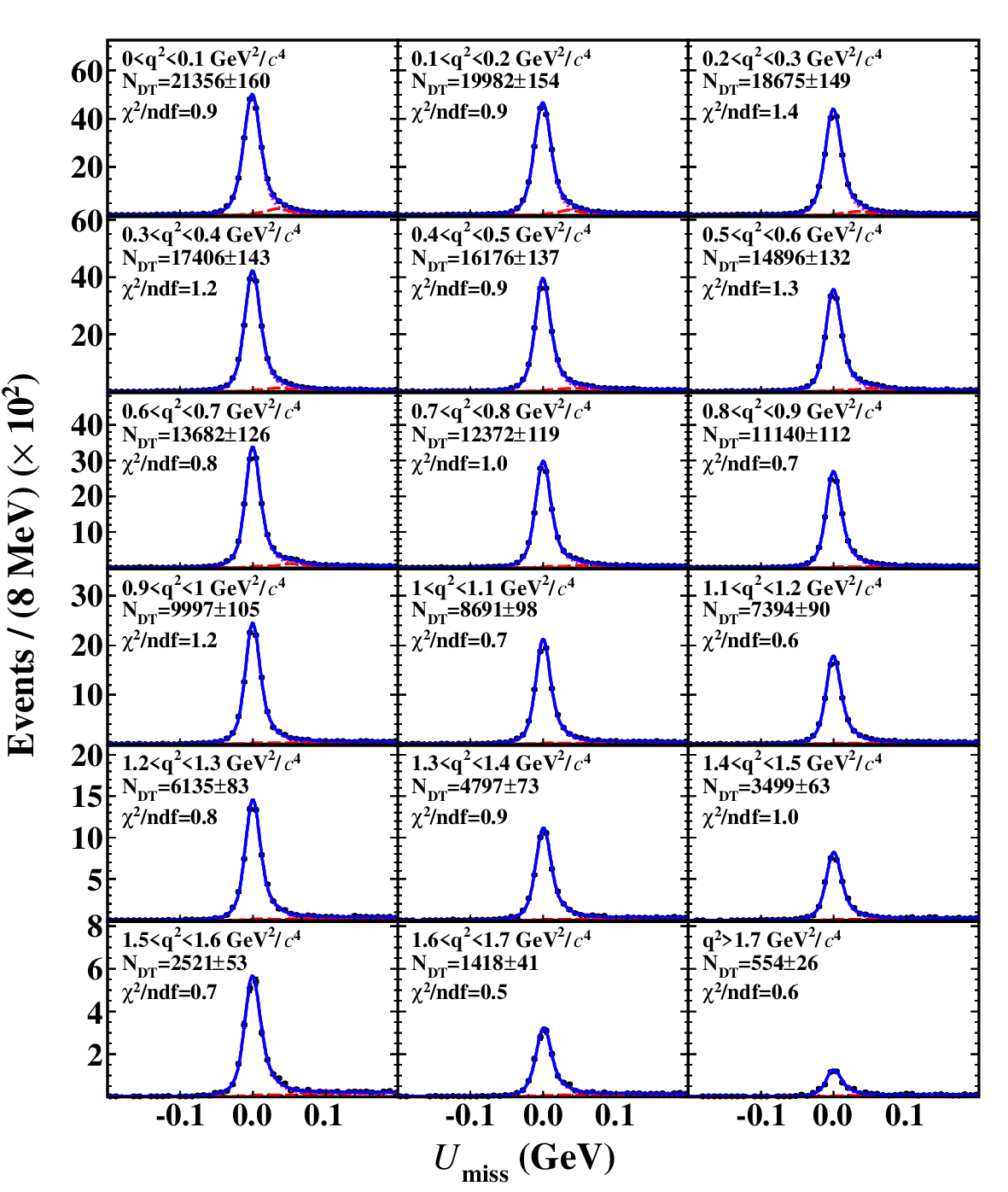}}
\caption{Fits to the $U_{\rm miss}$ distributions of the accepted $\kenu$ candidates in different $q^2$ bins.
The points with error bars are data. The blue solid curves are the fit results.
The violet dotted curves are the signal shapes, and
the red dashed curves are the fitted combinatorial background shapes.
\label{kenu_umissq2}
}
\end{center}
\end{figure*}

For each signal decay, the signal yield observed in each reconstructed
$q^{2}$ interval is obtained from a fit to the $U_{\rm miss}$
distribution. The fitting method is the same as mentioned in Section~\ref{bf_result}.
Figure~\ref{kenu_umissq2} shows the results of the fits
to the $U_{\rm miss}$ distributions in reconstructed $q^{2}$ intervals
for $\kenu$ semileptonic $D$ decays.  Similar figures for $\kmunu$,
$\ksenu$, and $\ksmunu$ decays are available in Figs. 7$-$9 in \hyperref[appendix]{Appendix}.

Table~\ref{tab:kenu_decayrate} lists the $q^{2}$ ranges, the fitted
numbers of observed DT events ($N_{\rm DT}$), the numbers of produced
events ($N_{\rm prd}$) calculated by the weighted efficiency matrix
and the decay rates ($\Delta\Gamma$) of $\kenu$ semileptonic $D$
decays in individual $q^2$ intervals. Similar tables for $\kmunu$,
$\ksenu$, and $\ksmunu$ decays are available in Tables 13$-$15 in \hyperref[appendix]{Appendix}.

\begin{table}[htbp]
\caption{ The observed yields ($N_{\rm DT}^i$), the produced yields ($N_{\rm
    prd}^i$) and the partial decay rates ($\Delta\Gamma$)
  of $\kenu$ in different $q^2$ intervals of data, where the
  uncertainties are statistical only. \label{tab:kenu_decayrate}}
\centering \resizebox{0.49\textwidth}{!}{
\begin{tabular}{cccc}
\hline
\hline
$q^2~({\rm GeV^2}/c^4)$&$N_{\rm DT}^i$&$N_{\rm prd}^i$&$\Delta\Gamma~({\rm ns^{-1}})$\\ \hline
$(0.00,0.10)$&  $21356\pm160$&$29580\pm236$&$9.100\pm0.073$ \\
$(0.10,0.20)$&  $19982\pm154$&$28248\pm247$&$8.690\pm0.076$ \\
$(0.20,0.30)$&  $18675\pm149$&$26707\pm249$&$8.216\pm0.076$ \\
$(0.30,0.40)$&  $17406\pm143$&$25180\pm244$&$7.746\pm0.075$ \\
$(0.40,0.50)$&  $16176\pm137$&$23475\pm238$&$7.221\pm0.073$ \\
$(0.50,0.60)$&  $14896\pm132$&$21685\pm231$&$6.671\pm0.071$ \\
$(0.60,0.70)$&  $13682\pm126$&$20003\pm222$&$6.153\pm0.068$ \\
$(0.70,0.80)$&  $12372\pm119$&$18119\pm211$&$5.574\pm0.065$ \\
$(0.80,0.90)$&  $11140\pm112$&$16357\pm198$&$5.032\pm0.061$ \\
$(0.90,1.00)$&  $9997\pm105$&$14719\pm187$&$4.528\pm0.057$ \\
$(1.00,1.10)$&  $8691\pm98$&$13040\pm176$&$4.011\pm0.054$ \\
$(1.10,1.20)$&  $7394\pm90$&$11091\pm162$&$3.412\pm0.050$ \\
$(1.20,1.30)$&  $6135\pm83$&$9459\pm150$&$2.910\pm0.046$ \\
$(1.30,1.40)$&  $4797\pm73$&$7644\pm136$&$2.352\pm0.042$ \\
$(1.40,1.50)$&  $3499\pm63$&$5627\pm118$&$1.731\pm0.036$ \\
$(1.50,1.60)$&  $2521\pm53$&$4356\pm105$&$1.340\pm0.032$ \\
$(1.60,1.70)$&  $1418\pm41$&$2621\pm86$&$0.806\pm0.026$ \\
$(1.70,1.88)$&  $554\pm26$&$1378\pm72$&$0.424\pm0.022$ \\
\hline
\hline
\end{tabular}
	}
\end{table}

\subsection{Construction of $\chi^2$ and statistical covariance matrices}

To determine the hadronic transition form factor and $|V_{cs}|$, a
least $\chi^{2}$ method is used to fit the partial decay rates of the
different signal decays.  Considering the correlations of the measured
partial decay rates ($\Delta\Gamma_i^{\rm msr}$) among different
$q^{2}$ intervals, the $\chi^{2}$ is given by
\begin{equation}\label{eq:chi}
	\chi^{2} = \sum_{i,j=1}^{N_{\mathrm{intervals}}}\left(\Delta\Gamma_{i}^{\mathrm{msr}}-\Delta\Gamma_{i}^{\mathrm{th}}\right) (C^{-1})_{ij}\left(\Delta\Gamma_{j}^{\mathrm{msr}}-\Delta\Gamma_{j}^{\mathrm{th}}\right),
\end{equation}
where $\Delta\Gamma_i^{\rm th}$ is the theoretically expected decay
rate in the $i$-th interval, $(C^{-1})_{ij}$ is the element of the
inverse covariance matrix of the measured partial decay rates and
is given by $C_{ij} = C_{ij}^{\mathrm{stat}}+C_{ij}^{\mathrm{syst}}$.
Here, $C_{ij}^{\mathrm{stat}}$ and $C_{ij}^{\mathrm{syst}}$ are
elements of the statistical and systematic covariance matrices,
respectively.  The elements of the statistical covariance matrix are
defined as
\begin{equation}
	C_{ij}^{\rm stat} =\left (\frac{1}{\tau_{D}N_{\mathrm{ST}}^{\rm tot}}\right)^{2}\sum_{\alpha}(\varepsilon^{-1})_{i\alpha}(\varepsilon^{-1})_{j\alpha}\left(\sigma\left(N_{\mathrm{DT}}^{\alpha}\right)\right)^{2},
\end{equation}
where $\sigma(N_{\mathrm{DT}}^{\alpha})$ is the statistical uncertainty of the signal yield observed in the $\alpha$-th interval.
The elements of the statistical covariance density matrices of
$\kenu$, $\kmunu$, $\koenu$ and $\komunu$ decays are presented in Tables 16$-$19 in \hyperref[appendix]{Appendix}.

\subsection{Systematic uncertainties on partial decay rates}

The sources of systematic uncertainties are discussed below.

\paragraph{\boldmath \bf ST $\bar D$ yields}

The systematic uncertainties associated with the number of $\bar D^0(D^-)$ tags are fully correlated across $q^2$ intervals.
The element of the related systematic covariance matrix is calculated by
\begin{equation}
C_{ij}^{\mathrm{syst}}\left(N_{\rm ST}\right)=\Delta\Gamma_{i}\Delta\Gamma_{j}\left(\frac{\sigma\left(N_{\rm ST}\right)}{N_{\rm ST}}\right)^2,
\end{equation}
where $\sigma(N_{\rm ST})/N_{\rm ST}$ is the relative uncertainty on the number of $\bar D^0(D^-)$ tags.

\paragraph{\boldmath \bf $D$ lifetime}

The systematic uncertainties associated with the $D$ lifetime are
fully correlated across the $q^2$ intervals. The element of the
related systematic covariance matrix is calculated by
\begin{equation}
C_{ij}^{\mathrm{syst}}\left(\tau_{D}\right)=\sigma\left(\Delta\Gamma_{i}\right)\sigma\left(\Delta\Gamma_{j}\right),
\end{equation}
where $\sigma(\Delta\Gamma_{i})=\sigma \tau_{D}\cdot\Delta\Gamma_{i}$ and $\sigma \tau_{D}$ is the uncertainty on the $D$ lifetime~\cite{pdg2022}.

\paragraph{\bf MC statistics}

The elements of the covariance matrix which accounts for the systematic
uncertainties and correlations between the $q^{2}$ intervals are calculated by
\begin{equation}
\begin{array}{l}
\displaystyle C_{ij}^{\mathrm{syst}}\left(\rm MC^{stat}\right)=\left(\frac{1}{\tau_{D}N_{\mathrm{ST}}}\right)^{2}\\
\displaystyle\times \sum_{\alpha\beta}N_{\mathrm{DT}}^{\alpha}N_{\mathrm{DT}}^{\beta}\mathrm{Cov}\left(\left(\varepsilon^{-1}\right)_{i\alpha},\left(\varepsilon^{-1}\right)_{j\beta}\right),
\end{array}
\end{equation}
where $N_{\mathrm{DT}}^{\alpha(\beta)}$ is the signal yield observed in the interval $\alpha(\beta)$, and the covariances of the inverse efficiency matrix elements are given by
\begin{equation}
\begin{array}{l}
\mathrm{Cov}\left(\left(\varepsilon^{-1}\right)_{i\alpha},\left(\varepsilon^{-1}\right)_{j\beta}\right)=\\ \sum\limits_{mn}\left(\left(\varepsilon^{-1}\right)_{im}\left(\varepsilon^{-1}\right)_{jm}\right)\left[\sigma\left(\varepsilon_{mn}\right)\right]^2\left(\left(\varepsilon^{-1}\right)_{\alpha n}\left(\varepsilon^{-1}\right)_{\beta n}\right).
\end{array}
\end{equation}

\paragraph{\bf MC model}

To estimate the uncertainty from the MC model, we vary the parameters
of the two-parameter series expansion model by $\pm1\sigma$.  The difference
between the alternative and nominal efficiencies is taken as the
systematic uncertainty for each signal decay.  The element of the
covariance matrix is defined as
\begin{equation}
C_{ij}^{\mathrm{syst}}\left (\mathrm{MC~model}\right )=\delta\left(\Delta\Gamma_{i}\right)\delta\left(\Delta\Gamma_{j}\right),
\end{equation}
	where $\delta(\Delta\Gamma_{i})$ denotes the change of the partial decay rate in the $i$-th $q^{2}$ interval.

\paragraph{\bf Tracking, PID}

The systematic uncertainties associated with the $e^+$ or $\mu^+$
tracking and PID efficiencies, and $K^-$ tracking and PID efficiencies
are estimated by varying the corresponding correction factors for
efficiencies within $\pm 1\sigma$. Using the new efficiency matrix,
the element of the corresponding systematic covariance matrix is
calculated by
\begin{equation}
C_{ij}^{\mathrm{syst}}\left (\mathrm{Tracking,~PID}\right)=\delta\left(\Delta\Gamma_{i}\right)\delta\left(\Delta\Gamma_{j}\right),
\end{equation}
where $\delta(\Delta\Gamma_{i})$ denotes the change of the partial decay rate in the $i$-th $q^{2}$ interval.

\paragraph{\boldmath \bf $U_{\rm miss}$ fit}

The systematic covariance matrix arising from the uncertainty in the $U_{\rm miss}$ fit has elements
\begin{equation}
C_{ij}^{\mathrm{syst}}\left(U_{\rm miss}~\mathrm{fit}\right)=\left(\frac{1}{\tau_{D}N_{\mathrm{ST}}^{\rm tot}}\right)^{2}\sum_{\alpha}\varepsilon_{i\alpha}^{-1}\varepsilon_{j\alpha}^{-1}\left(\sigma_{\alpha}^{\mathrm{Fit}}\right)^{2},
\end{equation}
where $\sigma_{\alpha}^{\mathrm{Fit}}$ is the systematic uncertainty
of the signal yield observed in the interval $\alpha$ obtained by
varying the background shape in the $U_{\rm miss}$ fit as described in Section~\ref{section_b}.

\paragraph{\bf Remaining uncertainties}

The remaining systematic uncertainties, include the $E_{\rm
  extra~\gamma}^{\rm max}$ and $N_{\rm extra}^{\rm trk}$ requirements,
$K^0_S$ reconstruction, and quoted branching fractions, are assumed to
be fully correlated across $q^{2}$ intervals, and the element of the
corresponding systematic covariance matrix is calculated by
\begin{equation}
C_{ij}^{\mathrm{syst}}=\sigma\left(\Delta\Gamma_{i}\right)\sigma\left(\Delta\Gamma_{j}\right),
\end{equation}
where $\sigma(\Delta\Gamma_{i})=\sigma_{\rm
  syst}\cdot\Delta\Gamma_{i}$.  The systematic uncertainties $\sigma_{\rm
  syst}$ on $\kenu$ semileptonic $D$ decays in different $q^2$
intervals are shown in Table~\ref{tab:kenu_sysq2}, and those of
$\kmunu$, $\ksenu$, and $\ksmunu$ decays are available in Tables 20$-$22 in \hyperref[appendix]{Appendix}, as well as the elements of the systematic covariance density
matrices for all signal decays.

\begin{table*}[htbp]
\caption{The systematic uncertainties (in \%) of the measured decay rates of $\kenu$ in different
$q^2$ bins.
\label{tab:kenu_sysq2}}
\centering
\begin{tabular}{ccccccccccccccccccc}
\hline
\hline
$i$-th $q^2$ bin&1&2&3&4&5&6&7&8&9&10&11&12&13&14&15&16&17&18\\ \hline
$N_{\rm ST}$&0.30&0.30&0.30&0.30&0.30&0.30&0.30&0.30&0.30&0.30&0.30&0.30&0.30&0.30&0.30&0.30&0.30&0.30\\
$D^0$ lifetime&0.24&0.24&0.24&0.24&0.24&0.24&0.24&0.24&0.24&0.24&0.24&0.24&0.24&0.24&0.24&0.24&0.24&0.24\\
MC statistics&0.14&0.15&0.15&0.16&0.17&0.17&0.18&0.19&0.20&0.21&0.23&0.25&0.28&0.31&0.36&0.45&0.60&0.99\\
MC model&0.21&0.19&0.56&0.28&0.06&0.46&1.50&0.25&1.58&0.31&0.44&0.04&0.57&1.33&0.31&0.29&3.63&1.04\\
$K^-$ tracking&0.10&0.10&0.10&0.10&0.10&0.10&0.10&0.10&0.10&0.10&0.10&0.11&0.11&0.13&0.13&0.18&0.23&0.43\\
$K^-$ PID&0.10&0.10&0.10&0.10&0.10&0.10&0.10&0.10&0.10&0.10&0.10&0.10&0.10&0.10&0.10&0.10&0.10&0.13\\
$e^+$ tracking&0.10&0.10&0.10&0.10&0.10&0.10&0.10&0.10&0.10&0.10&0.10&0.10&0.10&0.10&0.10&0.10&0.10&0.10\\
$e^+$ PID&0.10&0.10&0.10&0.10&0.10&0.10&0.10&0.10&0.10&0.10&0.10&0.10&0.10&0.10&0.10&0.10&0.10&0.10\\
$U_{\rm miss}$ fit&0.18&0.16&0.18&0.16&0.15&0.17&0.15&0.16&0.10&0.11&0.10&0.09&0.09&0.11&0.08&0.14&0.18&0.34\\
$E_{\rm extra~\gamma}^{\rm max}$ and $N_{\rm extra}^{\rm trk}$ cut&0.10&0.10&0.10&0.10&0.10&0.10&0.10&0.10&0.10&0.10&0.10&0.10&0.10&0.10&0.10&0.10&0.10&0.10\\

\hline
Total&0.54&0.53&0.75&0.57&0.50&0.68&1.58&0.57&1.65&0.59&0.67&0.52&0.78&1.44&0.66&0.73&3.71&1.60\\
\hline
\hline
\end{tabular}
\end{table*}

\subsection{Results based on individual fits}

With the statistical and systematic covariance matrices described previously, we fit individually the
partial decay rates of $\kenu$, $\kmunu$, $\ksenu$, and $\ksmunu$
to obtain the fit parameters $f_+^K(0)|V_{cs}|$ and $r_1(t_0)$ from Eq.~(\ref{ffk_function2}). The statistical
uncertainties on the fit parameters are taken from the fit with the
statistical covariance matrix, and the systematic uncertainties on the
fit parameters are obtained by calculating the quadrature difference
between the uncertainties of the fit parameters using the statistical
covariance matrix and the uncertainties using the combined statistical and
systematic covariance matrix.

\begin{figure*}[htbp]
\begin{center}
\includegraphics[width=0.80\textwidth]{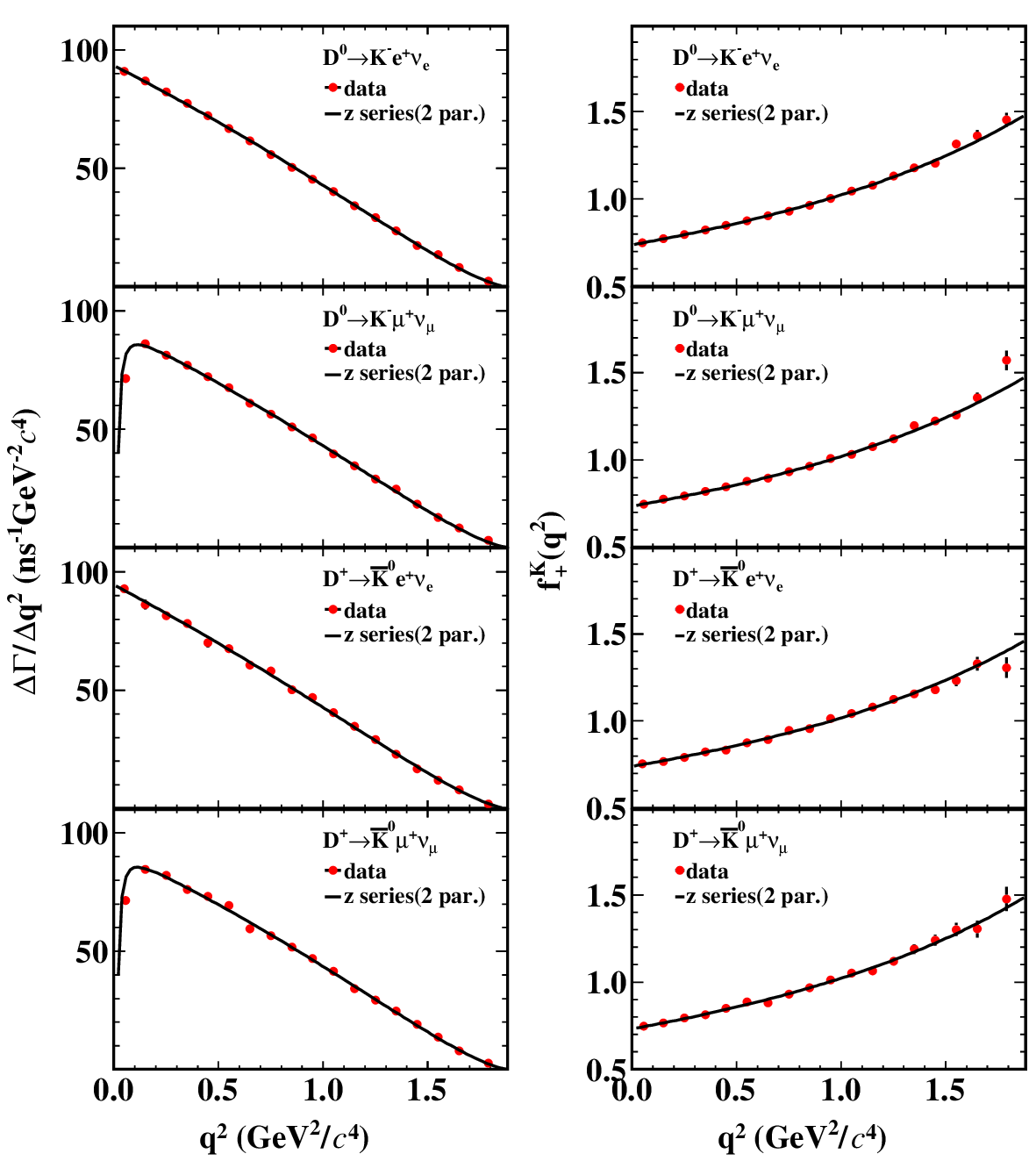}
\caption{(Left) Fits to the partial decay rates of $\klnu$ and (Right)
  projections of the form factor as functions of $q^2$, where the red
  points with error bars are the measured partial decay rates and the
  solid curves are the fits.
\label{fig:ff_klnu}
}
\end{center}
\end{figure*}

The sub-figures on the left of Fig.~\ref{fig:ff_klnu} show the
individual fit results of $\kenu$, $\kmunu$, $\ksenu$, and
$\ksmunu$. The sub-figures on the right of Fig.~\ref{fig:ff_klnu} show
the projections of the form factor fits as a function of $q^2$, where
the points with error bars show the measured values of the form
factor, which are obtained with
\begin{equation}
\begin{array}{l}
\displaystyle f_{+}^{\mathrm{data}}(q_i^{2})=
\displaystyle \sqrt{\frac{\left(\Delta\Gamma_i^{\mathrm{measured}}-B\right)\cdot|f_+\left(q_i^2\right)|^2}
{A}},\\
\end{array}
\end{equation}
with
\begin{equation}
\begin{array}{l}
\displaystyle A=\int_{q^{2}_{\mathrm{min}\left(i\right)}}^{q^{2}_{\mathrm{max}\left(i\right)}}\frac{G_{F}^{2}\left|V_{cs}\right|^{2}}{24\pi^{3}}\frac{\left(q^{2}-m^{2}_{\ell}\right)^2\left|\vec p_{K}\right|}{q^{4}m^{2}_{D}}\\
\times\left(1+\frac{m^{2}_{\ell}}{2q^{2}}\right)m^{2}_{D}\left|\vec p_{K}\right|^{2}\left|f_+(q^2)\right|^2dq^2,
\end{array}
\end{equation}
\begin{equation}
\begin{array}{l}
\displaystyle
B=\int_{q^{2}_{\mathrm{min}\left(i\right)}}^{q^{2}_{\mathrm{max}\left(i\right)}}\frac{G_{F}^{2}\left|V_{cs}\right|^{2}}{24\pi^{3}}\frac{\left(q^{2}-m^{2}_{\ell}\right)^2\left|\vec p_{K}\right|}{q^{4}m^{2}_{D}}\\
\times\frac{3m^{2}_{\ell}}{8q^{2}}\left(m^{2}_{D}-m^{2}_{K}\right)^{2}\left|f_0\left(q^2\right)\right|^2dq^2,
\end{array}
\end{equation}
where $q^{2}_{\mathrm{min}(i)}$ and $q^{2}_{\mathrm{max}(i)}$ are the low and high boundaries of the $i$-th $q^{2}$ bin. Both functions $f_{+}(q^{2})$ and $f_{+}(q_{i}^{2})$ are calculated using the two-parameter series expansion model.

The parameters obtained from the individual fits to the differential decay rates of $\kenu$, $\kmunu$, $\ksenu$, and $\ksmunu$ are listed in Table~\ref{tab:fit_par}.

\begin{table*}[htbp]
\centering
\caption{The parameters ($f^K_+(0)|V_{cs}|$, $r_1(t_0)$) of
  the hadronic form factors from the fits to the partial decay rates
  of the semileptonic decays, where the first and second uncertainties
  are statistical and systematic, respectively. The column labeled $\rho_{\rm 2par}$ gives the correlation
  coefficients of the two parameters, and ndf denotes the number of degrees of freedom.
\label{tab:fit_par}}
\scalebox{1.0}{
\begin{tabular}{|c|c|c|c|c|c|}
 \hline\hline
Case & Signal decay &$f^K_+(0)|V_{cs}|$ & $r_1(t_0)$ & $\rho_{\rm 2par}$ & $\chi^2/\rm ndf$\\
  \hline
\multirow{4}{*}{Individual fit}&$\kenu$   &$0.7179\pm0.0016\pm0.0017$&$-2.30\pm0.05\pm0.03$&0.48&16.3/16  \\
                               &$\kmunu$  &$0.7162\pm0.0022\pm0.0019$&$-2.28\pm0.08\pm0.02$&0.62&17.1/16  \\
                               &$\ksenu$  &$0.7207\pm0.0027\pm0.0035$&$-2.13\pm0.10\pm0.07$&0.29&13.0/16  \\
                               &$\ksmunu$ &$0.7124\pm0.0035\pm0.0032$&$-2.41\pm0.12\pm0.08$&0.45&10.6/16  \\  \hline
Simultaneous fit               &$\klnu$   &$0.7171\pm0.0011\pm0.0013$&$-2.28\pm0.04\pm0.02$&0.44&60.9/70  \\
\hline\hline
\end{tabular}
}
\end{table*}

\subsection{Results based on a simultaneous fit}

To consider the correlation effects in the measurements of the
hadronic form factor among the four signal decays, we perform a
simultaneous fit to the partial decay rates of $\kenu$, $\kmunu$,
$\ksenu$, and $\ksmunu$ to obtain the product $\ffK|V_{cs}|$ and $r_1(t_0)$.

In the simultaneous fit, the values of
$f_+^K(0)|V_{cs}|$ and $r_1(t_0)$ are shared among the four signal decays.
We still use the least $\chi^{2}$ method from Eq.~(\ref{eq:chi}) to obtain the fit parameters.
The $\Delta\Gamma_i$ for these four semileptonic
decay modes are combined into one vector with 72 components and the
elements of the covariance matrix for the combined $\Delta\Gamma_i$ are
redefined as $C_{ij}=C^{\rm stat}_{ij}+C^{\rm csyst}_{ij}+C^{\rm
  usyst}_{ij}$, $(i,j=1,2,3,...,71,72)$, where $C^{\rm stat}_{ij}$ is
the element of statistical covariance matrix, which is diagonal in
blocks, {\it i.e.}
$$C^{\rm stat}=
\begin{pmatrix}
A&0&0&0\\
0&B&0&0\\
0&0&C&0\\
0&0&0&D
\end{pmatrix}.
$$
Here $A$, $B$, $C$, and $D$ are the statistical covariance matrices obtained from each signal channel.
The element of the correlated systematic covariance matrix is
\begin{equation}
	\label{eq:csys_matrix}
	C^{\mathrm{csyst}}_{ij}=\delta(\Delta\Gamma_{i})\delta(\Delta\Gamma_{j}).
\end{equation}
The uncorrelated systematic covariance matrix is defined in blocks as
$$C^{\rm usyst}=
\begin{pmatrix}
a&0&0&0\\
0&b&0&0\\
0&0&c&0\\
0&0&0&d
\end{pmatrix}
,
$$
where $a$, $b$, $c$, and $d$ are the uncorrelated systematic covariance matrices obtained from each signal channel.

Then, the elements of covariance density matrix for the simultaneous fit are available in Tables 27$-$30 in \hyperref[appendix]{Appendix}.

\begin{figure*}[htbp]
\begin{center}
\includegraphics[width=1.0\textwidth]{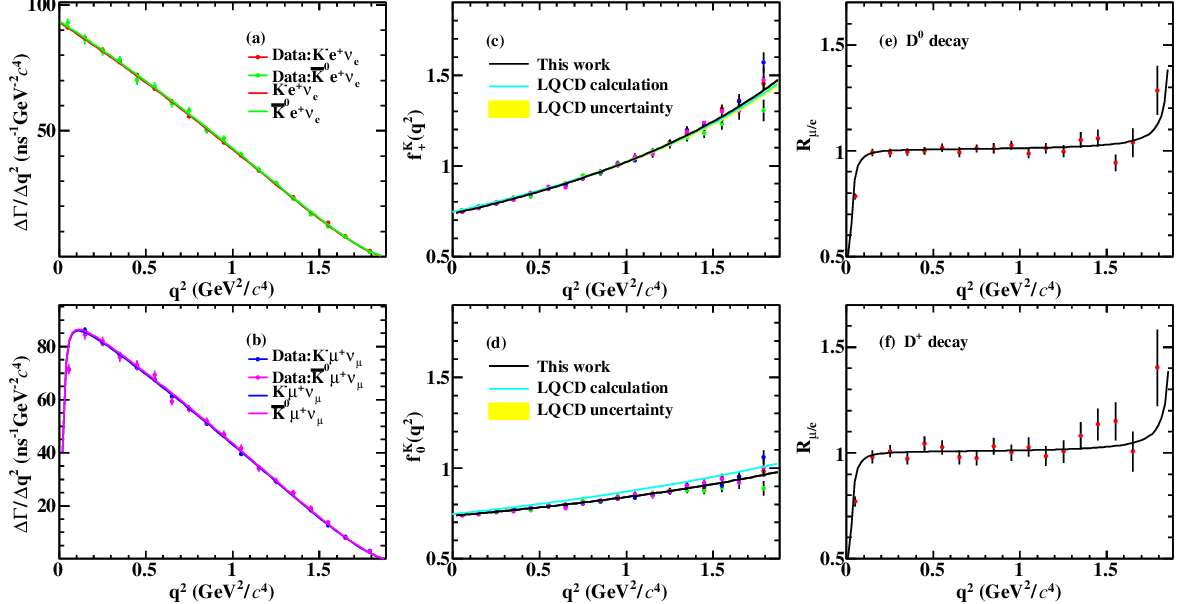}
\caption{(a)(b) Simultaneous fit to partial decay rates of
  $D^0(D^+)\to \bar K\ell^+\nu_\ell$. (c)(d) Projections of
  $f^{K}_+(q^2)$ and $f^{K}_0(q^2)$ as functions of $q^2$ of
  $D^0(D^+)\to \bar K\ell^+\nu_\ell$. (e)(f) The ratio of differential
  decay rates of $\kmunu$ over $\kenu$ and the ratio of differential
  decay rates of $\ksmunu$ over $\ksenu$ in each $q^2$ bin. The dots
  with error bars are data, and the solid lines are the results with the
  parameters of the simultaneous fit.}
\label{ff_klnu}
\end{center}
\end{figure*}

With the modified $\Delta\Gamma_i$ and $C_{ij}$, we do the simultaneous fit to the partial decay rates of $\klnu$, which is shown in Fig.~\ref{ff_klnu}.
The fitted parameters are  $f^K_+(0)|V_{cs}|=0.7171\pm0.0011\pm0.0013$ and $r_1(t_0)=-2.28\pm0.04\pm0.02$, which are summarized in Table~\ref{tab:fit_par}.

\section{Summary}

In summary, by analyzing 7.93 fb$^{-1}$ of $e^+e^-$ collision data
collected at $\sqrt{s}=3.773$~GeV with the BESIII detector, improved
measurements of the semileptonic decays $\klnu$ are performed. The
absolute branching fractions of $\kenu$, $\kmunu$, $\koenu$, and
$\komunu$ are determined to be $(3.521\pm0.009_{\rm
stat.}\pm0.016_{\rm syst.}) \%$, $(3.419\pm0.011_{\rm
stat.}\pm0.016_{\rm syst.}) \%$, $(8.864\pm0.039_{\rm
stat.}\pm0.082_{\rm syst.}) \%$ and $(8.665\pm0.046_{\rm
stat.}\pm0.084_{\rm syst.}) \%$, respectively. Combining the branching
fractions of semielectronic and semimuonic decays, we obtain the
ratios of the two branching fractions $\frac{\mathcal
B_{\kmunu}}{\mathcal B_{\kenu}}=0.971\pm0.004_{\rm stat.}\pm0.006_{\rm
syst.}$ and $\frac{\mathcal B_{\ksmunu}}{\mathcal
B_{\ksenu}}=0.978\pm0.007_{\rm stat.}\pm0.013_{\rm syst.}$, which are
consistent with the theoretical calculation
$0.975\pm0.001$~\cite{Riggio:2017zwh}. Our measurements support lepton
flavor universality.

\begin{figure*}[htp]\centering
\setlength{\abovecaptionskip}{-2pt}
\setlength{\belowcaptionskip}{-3pt}
\includegraphics[width=0.75\textwidth]{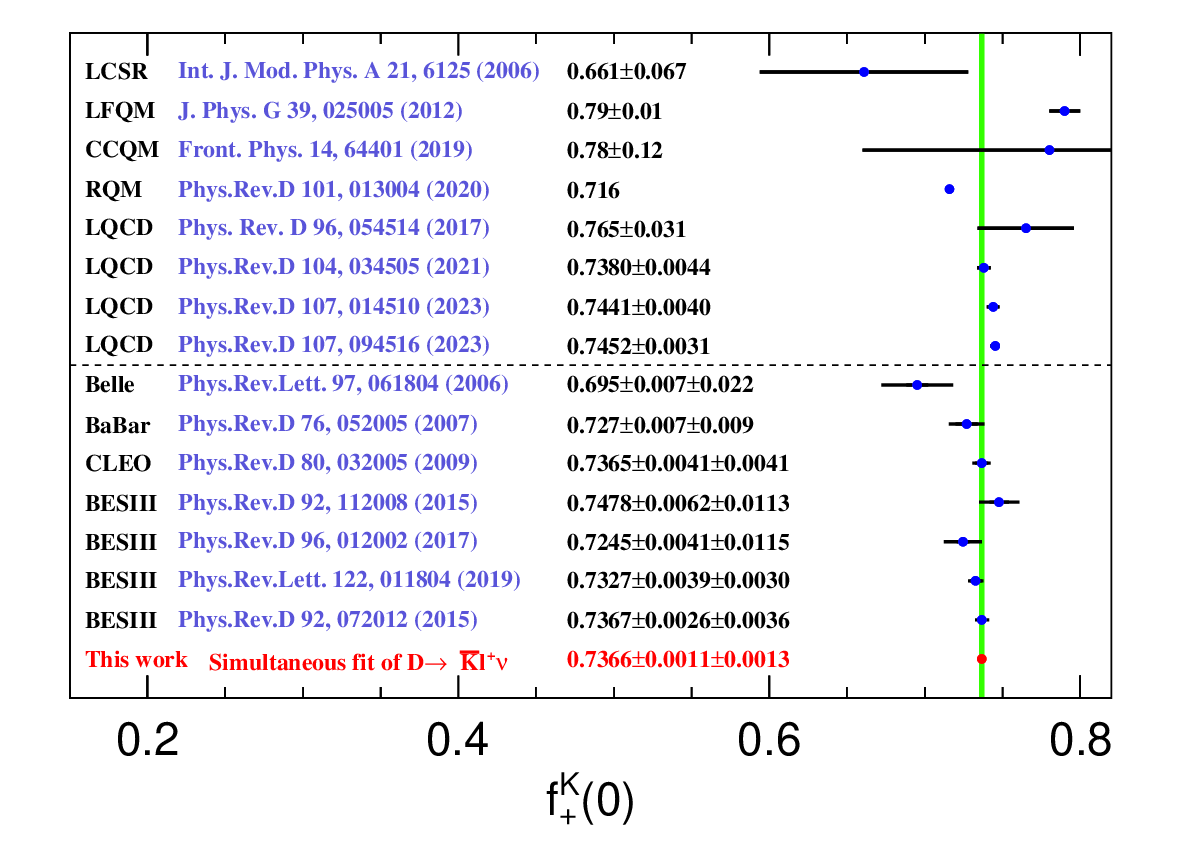}
\caption{Comparisons of the form factor $f_+^{K}(0)$ measured in this work with the theoretical and experimental calculations.
The first and second uncertainties are statistical and systematic, respectively.  The green band corresponds to the $\pm1\sigma$ limit of the form factor calculated in this work.
\label{compare_ff_klnu}
}
\end{figure*}

From the simultaneous fit to the partial decay rates of $\kenu$,
$\kmunu$, $\koenu$, and $\komunu$, the product of the hadronic form
factor $\ffK$ and the modulus of the CKM matrix element $|V_{cs}|$ are
determined to be $\ffK|V_{cs}|=0.7171\pm0.0011_{\rm
stat.}\pm0.0013_{\rm syst.}$. Taking the value of $|V_{cs}| = 0.97349\pm0.00016$ given by
the PDG~\cite{pdg2022} as input, we obtain the hadronic form factor
$\ffK=0.7366\pm0.0011_{\rm stat.}\pm0.0013_{\rm syst.}$.  Conversely,
using the $\ffK$ calculated in LQCD~\cite{FermilabLattice:2022gku}, we obtain
$|V_{cs}|=0.9623\pm0.0015_{\rm stat.}\pm0.0017_{\rm
syst.}\pm0.0040_{\rm LQCD}$.  The comparison of the $\ffK$ value obtained in
this work with the theoretical and experimental calculations is shown in Fig.~\ref{compare_ff_klnu}.
For the experimental calculations, the values of $\ffK$ are updated by using the latest value of $|V_{cs}|$ as mentioned before.
The hadronic form factor $\ffK$ measured in this work supersedes the previous BESIII results~\cite{BESIII:2015tql,BESIII:2018ccy,BESIII:2017ylw,BESIII:2015jmz}, with the better precision.
This is important to test different models and help to improve the precision of theoretical calculations.

\begin{acknowledgments}

The BESIII Collaboration thanks the staff of BEPCII and the IHEP computing center for their strong support. This work is supported in part by National Key R\&D Program of China under Contracts Nos. 2020YFA0406000, 2023YFA1606400, 2020YFA0406300; National Natural Science Foundation of China (NSFC) under Contracts Nos. 11635010, 11735014, 11935015, 11935016, 11935018, 12025502, 12035009, 12035013, 12061131003, 12192260, 12192261, 12192262, 12192263, 12192264, 12192265, 12221005, 12225509, 12235017, 12361141819; the Chinese Academy of Sciences (CAS) Large-Scale Scientific Facility Program; the CAS Center for Excellence in Particle Physics (CCEPP); Joint Large-Scale Scientific Facility Funds of the NSFC and CAS under Contract No. U1832207; 100 Talents Program of CAS; The Institute of Nuclear and Particle Physics (INPAC) and Shanghai Key Laboratory for Particle Physics and Cosmology; German Research Foundation DFG under Contracts Nos. 455635585, FOR5327, GRK 2149; Istituto Nazionale di Fisica Nucleare, Italy; Ministry of Development of Turkey under Contract No. DPT2006K-120470; National Research Foundation of Korea under Contract No. NRF-2022R1A2C1092335; National Science and Technology fund of Mongolia; National Science Research and Innovation Fund (NSRF) via the Program Management Unit for Human Resources \& Institutional Development, Research and Innovation of Thailand under Contracts Nos. B16F640076, B50G670107; Polish National Science Centre under Contract No. 2019/35/O/ST2/02907; The Swedish Research Council; U. S. Department of Energy under Contract No. DE-FG02-05ER41374.

\end{acknowledgments}

\newpage
\onecolumngrid
\appendix
\section*{Appendix}
\label{appendix}

Tables~\ref{kenu_effmatrix}, \ref{kmunu_effmatrix}, \ref{ksenu_effmatrix}, and \ref{ksmunu_effmatrix} report the elements of the weighted efficiency matrices for $\kenu$, $\kmunu$, $\koenu$, and $\komunu$, respectively.

\begin{figure*}[htbp]
\begin{center}
\subfigure{\includegraphics[width=0.98\textwidth]{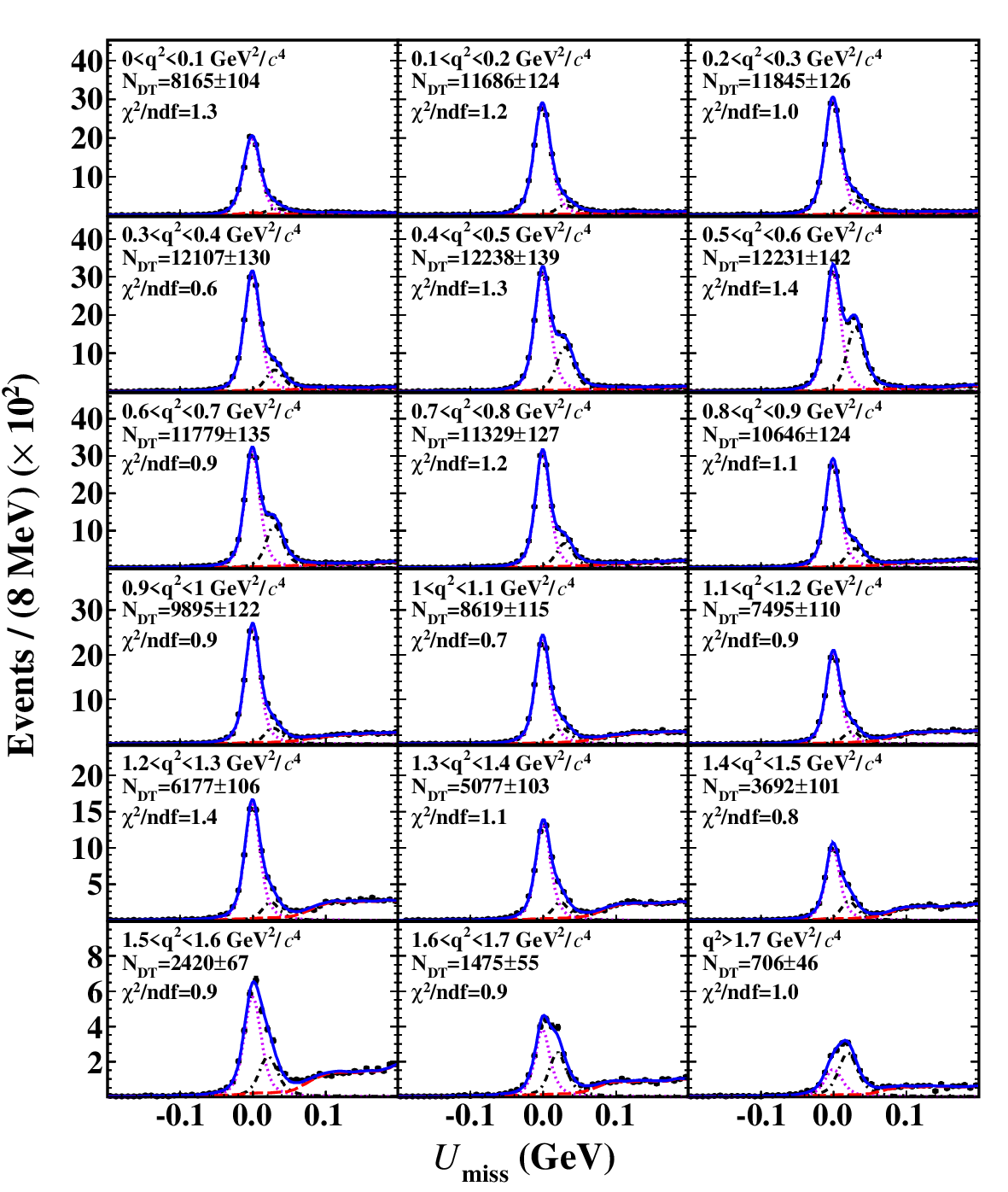}}
\caption{Fits to the $U_{\rm miss}$ distributions of the accepted $\kmunu$ candidates in different $q^2$ bins.
The points with  error bars are data. The blue solid curves are the fit results.
The violet dotted curves are the signal shapes.
The black dash-dotted curves are the peaking backgrounds.
The red dashed curves are the fitted combinatorial background shapes.
\label{kmunu_umissq2}
}
\end{center}
\end{figure*}

\begin{figure*}[htbp]
\begin{center}
\subfigure{\includegraphics[width=0.98\textwidth]{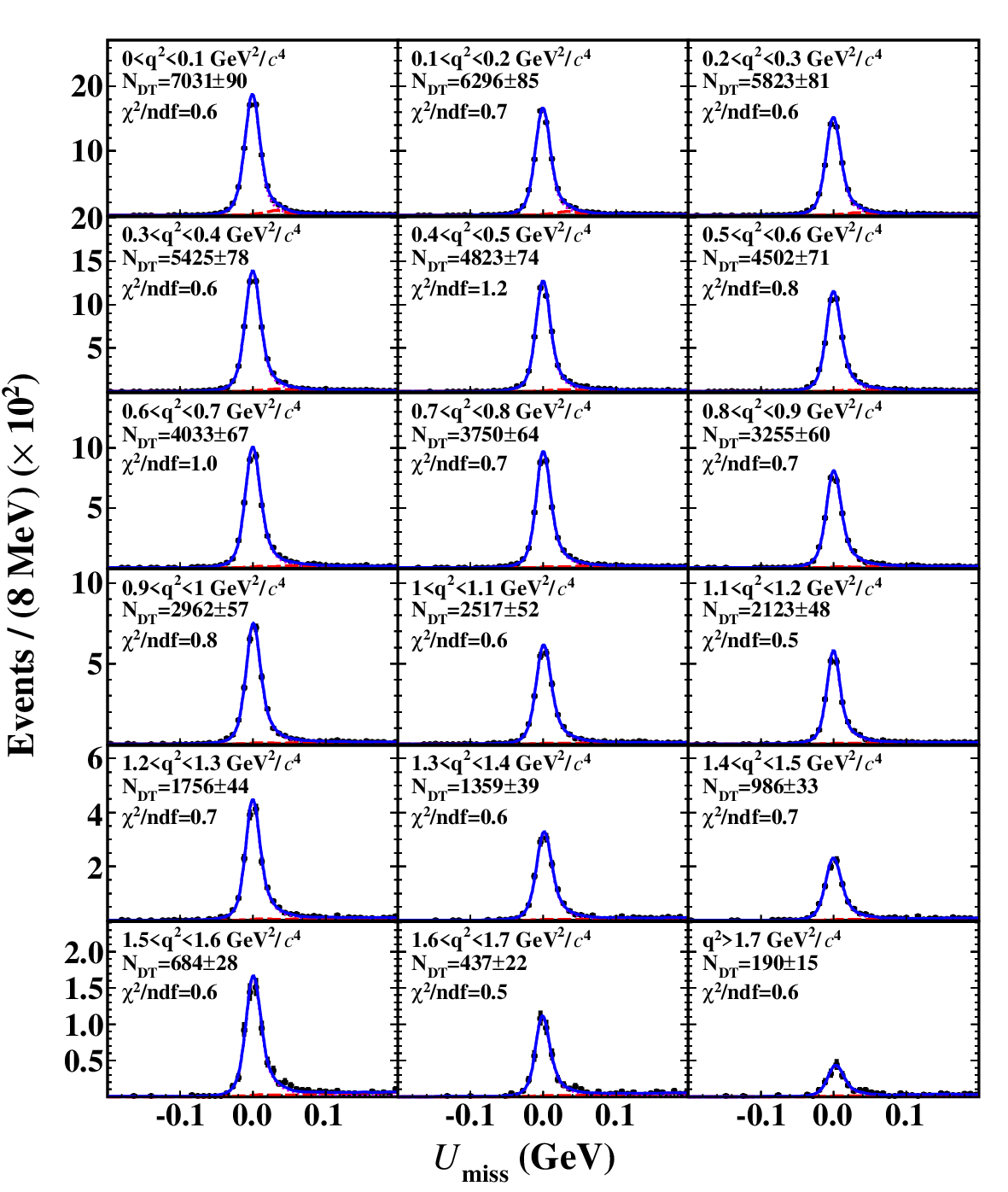}}
\caption{Fits to the $U_{\rm miss}$ distributions of the accepted $\ksenu$ candidates in different $q^2$ bins.
The points with error bars are data. The blue solid curves are the fit results.
The violet dotted curves are the signal shapes.
The red dashed curves are the fitted combinatorial background shapes.
\label{ksenu_umissq2}
}
\end{center}
\end{figure*}

\begin{figure*}[htbp]
\begin{center}
\subfigure{\includegraphics[width=0.98\textwidth]{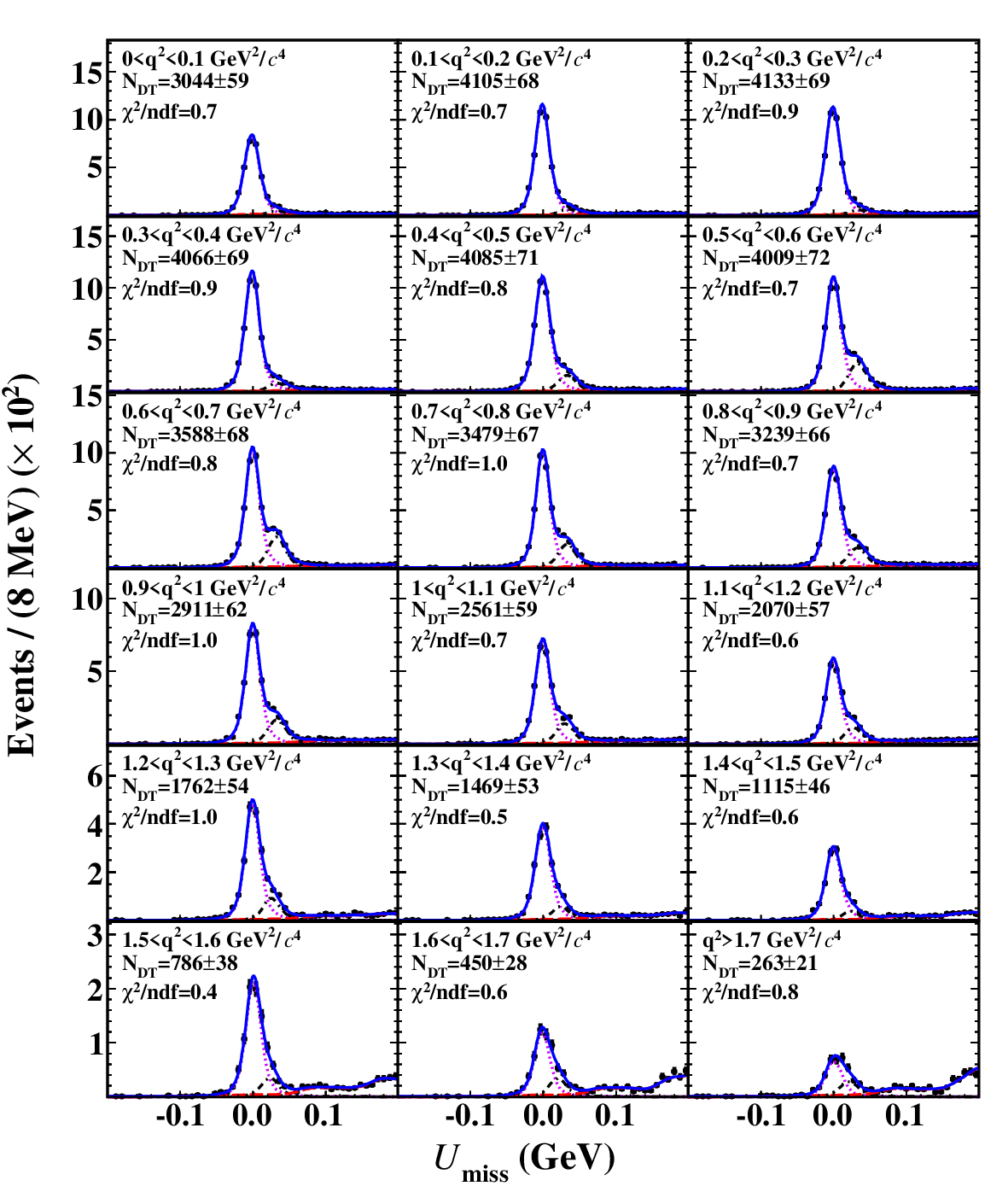}}
\caption{Fits to the $U_{\rm miss}$ distributions of the accepted $\ksmunu$ candidates in different $q^2$ bins.
The points with  error bars are data. The blue solid curves are the fit results.
The violet dotted curves are the signal shapes.
The black dash-dotted curves are the peaking backgrounds.
The red dashed curves are the fitted combinatorial background shapes. The bumps in the high $q^2$ region are mainly from the contributions of
$D^+\to K_S^0\pi^+\pi^0\pi^0$ and $D^+\to K_L^0K_S^0\pi^+$.
\label{ksmunu_umissq2}
}
\end{center}
\end{figure*}

Figures~\ref{kmunu_umissq2}, \ref{ksenu_umissq2}, and~\ref{ksmunu_umissq2} show the results of the fits to the $U_{\rm miss}$ distributions in the reconstructed $q^{2}$ intervals for $\kmunu$, $\koenu$, and $\komunu$, respectively.

Table~\ref{tab:kmunu_decayrate}, \ref{tab:ksenu_decayrate}, and \ref{tab:ksmunu_decayrate} list the $q^{2}$ ranges, the fitted numbers of observed DT events ($N_{\rm DT}$), the numbers of produced events ($N_{\rm prd}$) calculated by the weighted efficiency matrix and the decay rates ($\Delta\Gamma$) of $\kmunu$, $\koenu$, and $\komunu$ in individual $q^2$ intervals.

Tables~\ref{tab:kenu_statmatrix}, \ref{tab:kmunu_statmatrix}, \ref{tab:ksenu_statmatrix}, and \ref{tab:ksmunu_statmatrix} report the elements of the statistical covariance density matrices for $\kenu$, $\kmunu$, $\koenu$, and $\komunu$, respectively.

Table~\ref{tab:kmunu_sysq2}, \ref{tab:ksenu_sysq2}, and \ref{tab:ksmunu_sysq2} show the systematic uncertainties $\sigma_{\rm syst}$ of $\kmunu$, $\koenu$, and $\komunu$ in different $q^2$ intervals.

Tables~\ref{tab:kenu_sysmatrix}, \ref{tab:kmunu_sysmatrix}, \ref{tab:ksenu_sysmatrix} and \ref{tab:ksmunu_sysmatrix} report the elements of the systematic covariance density matrices for $\kenu$, $\kmunu$, $\koenu$, and $\komunu$, respectively.

Tables~\ref{klnumatrix_cov1},~\ref{klnumatrix_cov2},~\ref{klnumatrix_cov3}, and~\ref{klnumatrix_cov4} report the elements of the covariance density matrix $\rho_{ij}(i=0,1,2,...,71,72; j=0,1,2,...,71,72;)$ for the simultaneous fit.

\begin{table*}[htbp]
\caption{The weighted efficiency matrix (in \%) for $\kenu$. The $i$ denote the reconstructed bin, and
the $j$ represent the produced bin.
\label{kenu_effmatrix}}
\centering
\resizebox{1.0\textwidth}{!}{
\begin{tabular}{c|cccccccccccccccccc}
\hline
\hline
$\varepsilon_{ij}$&1&2&3&4&5&6&7&8&9&10&11&12&13&14&15&16&17&18\\ \hline
1&67.94&4.03&0.31&0.13&0.02&0.00&0.00&0.00&0.00&0.00&0.00&0.00&0.00&0.00&0.00&0.00&0.00&0.00\\
2&2.48&62.81&5.09&0.41&0.14&0.03&0.01&0.01&0.00&0.00&0.00&0.00&0.00&0.00&0.00&0.00&0.00&0.00\\
3&0.08&3.17&60.81&5.47&0.42&0.13&0.03&0.01&0.01&0.01&0.00&0.00&0.00&0.00&0.00&0.00&0.00&0.00\\
4&0.03&0.11&3.57&59.42&5.62&0.46&0.11&0.03&0.01&0.01&0.01&0.00&0.00&0.00&0.00&0.00&0.00&0.00\\
5&0.01&0.03&0.14&3.87&58.75&5.74&0.44&0.10&0.03&0.02&0.01&0.01&0.00&0.00&0.00&0.00&0.00&0.00\\
6&0.01&0.02&0.05&0.16&3.97&58.38&5.69&0.42&0.11&0.04&0.02&0.01&0.00&0.00&0.00&0.00&0.00&0.00\\
7&0.01&0.01&0.02&0.06&0.19&4.09&57.97&5.72&0.42&0.11&0.03&0.02&0.01&0.01&0.00&0.00&0.00&0.00\\
8&0.00&0.01&0.01&0.02&0.06&0.21&4.18&57.80&5.60&0.38&0.10&0.05&0.01&0.01&0.00&0.01&0.00&0.00\\
9&0.00&0.00&0.01&0.01&0.03&0.07&0.22&4.21&57.74&5.41&0.39&0.09&0.04&0.01&0.00&0.00&0.00&0.00\\
10&0.00&0.00&0.00&0.01&0.01&0.04&0.08&0.25&4.25&57.65&5.25&0.38&0.11&0.04&0.01&0.01&0.00&0.00\\
11&0.00&0.00&0.00&0.00&0.01&0.01&0.03&0.08&0.24&4.09&56.97&5.01&0.33&0.08&0.04&0.01&0.00&0.00\\
12&0.00&0.00&0.00&0.00&0.00&0.01&0.02&0.03&0.08&0.25&4.01&56.95&4.91&0.31&0.06&0.03&0.01&0.00\\
13&0.00&0.00&0.00&0.00&0.00&0.00&0.01&0.01&0.03&0.08&0.29&3.86&55.87&4.51&0.28&0.06&0.01&0.00\\
14&0.00&0.00&0.00&0.00&0.00&0.00&0.00&0.01&0.01&0.02&0.08&0.27&3.65&54.36&4.23&0.24&0.03&0.00\\
15&0.00&0.00&0.00&0.00&0.00&0.00&0.00&0.00&0.00&0.01&0.02&0.06&0.24&3.46&53.75&3.96&0.15&0.00\\
16&0.00&0.00&0.00&0.00&0.00&0.00&0.00&0.00&0.00&0.00&0.00&0.02&0.06&0.19&3.06&51.27&3.50&0.10\\
17&0.00&0.00&0.00&0.00&0.00&0.00&0.00&0.00&0.00&0.00&0.00&0.00&0.01&0.03&0.15&2.65&47.84&2.71\\
18&0.00&0.00&0.00&0.00&0.00&0.00&0.00&0.00&0.00&0.00&0.00&0.00&0.00&0.00&0.01&0.08&1.77&36.54\\
\hline
\hline
\end{tabular}
	}
\end{table*}

\begin{table*}[htbp]
\caption{The weighted efficiency matrix (in \%) for $\kmunu$. The $i$ denote the reconstructed bin, and
the $j$ represent the produced bin.
\label{kmunu_effmatrix}}
\centering
\resizebox{1.0\textwidth}{!}{
\begin{tabular}{c|cccccccccccccccccc}
\hline
\hline
$\varepsilon_{ij}$&1&2&3&4&5&6&7&8&9&10&11&12&13&14&15&16&17&18\\ \hline
1&38.04&1.14&0.02&0.00&0.00&0.00&0.00&0.00&0.00&0.00&0.00&0.00&0.00&0.00&0.00&0.00&0.00&0.00\\
2&1.52&38.87&1.79&0.03&0.01&0.00&0.00&0.00&0.00&0.00&0.00&0.00&0.00&0.00&0.00&0.00&0.00&0.00\\
3&0.04&1.71&40.76&2.28&0.05&0.01&0.00&0.00&0.00&0.00&0.00&0.00&0.00&0.00&0.00&0.00&0.00&0.00\\
4&0.01&0.05&2.17&43.37&2.71&0.08&0.02&0.01&0.00&0.00&0.00&0.00&0.00&0.00&0.00&0.00&0.00&0.00\\
5&0.01&0.02&0.08&2.61&46.18&3.14&0.13&0.03&0.02&0.01&0.00&0.00&0.00&0.00&0.00&0.00&0.00&0.00\\
6&0.01&0.01&0.03&0.10&2.99&49.00&3.46&0.15&0.05&0.02&0.01&0.00&0.00&0.00&0.00&0.00&0.00&0.00\\
7&0.00&0.01&0.02&0.03&0.14&3.33&51.75&3.70&0.18&0.06&0.03&0.01&0.00&0.00&0.00&0.00&0.00&0.00\\
8&0.00&0.00&0.01&0.02&0.06&0.18&3.65&53.83&3.88&0.21&0.06&0.02&0.01&0.01&0.00&0.00&0.00&0.00\\
9&0.00&0.00&0.00&0.01&0.02&0.05&0.18&3.80&55.97&3.93&0.23&0.08&0.04&0.02&0.00&0.00&0.00&0.00\\
10&0.00&0.00&0.00&0.01&0.01&0.03&0.07&0.23&3.85&57.24&3.97&0.27&0.08&0.03&0.01&0.01&0.00&0.00\\
11&0.00&0.00&0.00&0.00&0.01&0.02&0.03&0.07&0.24&3.97&58.22&3.82&0.23&0.07&0.03&0.01&0.00&0.00\\
12&0.00&0.00&0.00&0.00&0.01&0.01&0.01&0.03&0.07&0.25&3.90&58.46&3.73&0.23&0.08&0.02&0.00&0.00\\
13&0.00&0.00&0.00&0.00&0.00&0.00&0.01&0.01&0.03&0.08&0.28&3.76&57.18&3.58&0.20&0.05&0.01&0.00\\
14&0.00&0.00&0.00&0.00&0.00&0.00&0.00&0.00&0.01&0.02&0.07&0.28&3.47&56.01&3.29&0.17&0.03&0.01\\
15&0.00&0.00&0.00&0.00&0.00&0.00&0.00&0.00&0.00&0.01&0.02&0.06&0.25&3.26&54.77&3.09&0.12&0.02\\
16&0.00&0.00&0.00&0.00&0.00&0.00&0.00&0.00&0.00&0.00&0.01&0.01&0.05&0.19&2.95&52.16&2.87&0.05\\
17&0.00&0.00&0.00&0.00&0.00&0.00&0.00&0.00&0.00&0.00&0.00&0.00&0.01&0.03&0.14&2.42&48.84&2.01\\
18&0.00&0.00&0.00&0.00&0.00&0.00&0.00&0.00&0.00&0.00&0.00&0.00&0.00&0.00&0.02&0.07&1.70&37.02\\
\hline
\hline
\end{tabular}
	}
\end{table*}

\begin{table*}[htbp]
\caption{The weighted efficiency matrix (in \%) for $\ksenu$. The $i$ denote the reconstructed bin, and
the $j$ represent the produced bin.
\label{ksenu_effmatrix}}
\centering
\resizebox{1.0\textwidth}{!}{
\begin{tabular}{c|cccccccccccccccccc}
\hline
\hline
$\varepsilon_{ij}$&1&2&3&4&5&6&7&8&9&10&11&12&13&14&15&16&17&18\\ \hline
1&48.53&2.67&0.17&0.07&0.01&0.00&0.00&0.00&0.00&0.00&0.00&0.00&0.00&0.00&0.00&0.00&0.00&0.00\\
2&1.49&44.41&3.35&0.22&0.07&0.00&0.00&0.00&0.00&0.00&0.00&0.00&0.00&0.00&0.00&0.00&0.00&0.00\\
3&0.03&1.84&42.69&3.62&0.22&0.06&0.00&0.00&0.00&0.00&0.00&0.00&0.00&0.00&0.00&0.00&0.00&0.00\\
4&0.00&0.04&2.04&41.28&3.70&0.20&0.04&0.00&0.00&0.00&0.00&0.00&0.00&0.00&0.00&0.00&0.00&0.00\\
5&0.00&0.01&0.05&2.16&40.20&3.77&0.20&0.03&0.00&0.00&0.00&0.00&0.00&0.00&0.00&0.00&0.00&0.00\\
6&0.00&0.00&0.01&0.06&2.25&39.14&3.73&0.17&0.01&0.00&0.00&0.00&0.00&0.00&0.00&0.00&0.00&0.00\\
7&0.00&0.00&0.00&0.01&0.06&2.31&38.65&3.66&0.15&0.01&0.00&0.00&0.00&0.00&0.00&0.00&0.00&0.00\\
8&0.00&0.00&0.00&0.00&0.01&0.06&2.39&37.92&3.54&0.14&0.01&0.00&0.00&0.00&0.00&0.00&0.00&0.00\\
9&0.00&0.00&0.00&0.00&0.00&0.01&0.07&2.42&37.71&3.36&0.11&0.01&0.00&0.00&0.00&0.00&0.00&0.00\\
10&0.00&0.00&0.00&0.00&0.00&0.00&0.01&0.08&2.34&37.20&3.25&0.10&0.01&0.00&0.00&0.00&0.00&0.00\\
11&0.00&0.00&0.00&0.00&0.00&0.00&0.00&0.01&0.07&2.32&36.59&3.07&0.08&0.00&0.00&0.00&0.00&0.00\\
12&0.00&0.00&0.00&0.00&0.00&0.00&0.00&0.00&0.02&0.08&2.32&36.15&2.87&0.06&0.00&0.00&0.00&0.00\\
13&0.00&0.00&0.00&0.00&0.00&0.00&0.00&0.00&0.01&0.01&0.08&2.21&35.89&2.71&0.04&0.00&0.00&0.00\\
14&0.00&0.00&0.00&0.00&0.00&0.00&0.00&0.00&0.00&0.00&0.01&0.08&2.06&35.51&2.53&0.03&0.00&0.00\\
15&0.00&0.00&0.00&0.00&0.00&0.00&0.00&0.00&0.00&0.00&0.00&0.01&0.09&2.00&35.05&2.25&0.02&0.00\\
16&0.00&0.00&0.00&0.00&0.00&0.00&0.00&0.00&0.00&0.00&0.00&0.00&0.01&0.07&1.75&34.66&2.00&0.01\\
17&0.00&0.00&0.00&0.00&0.00&0.00&0.00&0.00&0.00&0.00&0.00&0.00&0.00&0.00&0.06&1.54&34.31&1.77\\
18&0.00&0.00&0.00&0.00&0.00&0.00&0.00&0.00&0.00&0.00&0.00&0.00&0.00&0.00&0.01&0.04&1.33&32.87\\
\hline
\hline
\end{tabular}
	}
\end{table*}

\begin{table*}[htbp]
\caption{The weighted efficiency matrix (in \%) for $\ksmunu$. The $i$ denote the reconstructed bin, and
the $j$ represent the produced bin.
\label{ksmunu_effmatrix}}
\centering
\resizebox{1.0\textwidth}{!}{
\begin{tabular}{c|cccccccccccccccccc}
\hline
\hline
$\varepsilon_{ij}$&1&2&3&4&5&6&7&8&9&10&11&12&13&14&15&16&17&18\\ \hline
1&31.24&0.90&0.00&0.00&0.00&0.00&0.00&0.00&0.00&0.00&0.00&0.00&0.00&0.00&0.00&0.00&0.01&0.01\\
2&1.05&30.73&1.36&0.01&0.00&0.00&0.00&0.00&0.00&0.00&0.00&0.00&0.00&0.00&0.01&0.01&0.01&0.00\\
3&0.02&1.12&31.34&1.67&0.01&0.00&0.00&0.00&0.00&0.00&0.00&0.00&0.00&0.00&0.00&0.00&0.01&0.01\\
4&0.00&0.02&1.42&32.64&2.01&0.03&0.00&0.00&0.00&0.00&0.00&0.00&0.00&0.00&0.00&0.00&0.00&0.01\\
5&0.00&0.00&0.02&1.62&33.99&2.16&0.02&0.00&0.00&0.00&0.00&0.00&0.00&0.00&0.00&0.00&0.00&0.00\\
6&0.00&0.00&0.00&0.03&1.86&35.12&2.31&0.02&0.00&0.00&0.00&0.00&0.00&0.00&0.00&0.01&0.02&0.01\\
7&0.00&0.00&0.00&0.00&0.04&2.01&36.06&2.44&0.04&0.00&0.00&0.01&0.00&0.00&0.01&0.01&0.03&0.02\\
8&0.00&0.00&0.00&0.00&0.01&0.04&2.19&36.89&2.50&0.03&0.01&0.00&0.00&0.01&0.01&0.01&0.01&0.01\\
9&0.00&0.00&0.00&0.00&0.01&0.01&0.05&2.28&37.36&2.52&0.04&0.01&0.00&0.00&0.00&0.01&0.01&0.00\\
10&0.00&0.00&0.00&0.00&0.00&0.00&0.01&0.07&2.24&37.14&2.44&0.02&0.01&0.01&0.01&0.01&0.00&0.00\\
11&0.00&0.00&0.00&0.00&0.00&0.01&0.01&0.02&0.06&2.26&37.01&2.34&0.03&0.01&0.00&0.01&0.01&0.00\\
12&0.00&0.00&0.00&0.00&0.00&0.00&0.00&0.01&0.02&0.07&2.21&36.36&2.11&0.03&0.01&0.01&0.00&0.00\\
13&0.00&0.00&0.00&0.00&0.01&0.01&0.00&0.01&0.00&0.01&0.07&2.12&36.27&2.08&0.02&0.01&0.00&0.00\\
14&0.00&0.00&0.00&0.00&0.01&0.00&0.00&0.01&0.01&0.01&0.01&0.08&2.04&36.00&1.91&0.01&0.00&0.00\\
15&0.00&0.01&0.00&0.00&0.00&0.00&0.01&0.01&0.01&0.01&0.02&0.02&0.06&1.93&35.32&1.78&0.01&0.00\\
16&0.00&0.00&0.00&0.00&0.01&0.00&0.01&0.01&0.00&0.01&0.01&0.01&0.01&0.05&1.69&34.99&1.61&0.00\\
17&0.00&0.01&0.00&0.00&0.01&0.01&0.01&0.01&0.01&0.00&0.00&0.00&0.00&0.01&0.05&1.56&34.29&1.28\\
18&0.01&0.01&0.01&0.01&0.01&0.01&0.00&0.00&0.00&0.00&0.00&0.00&0.00&0.00&0.01&0.03&1.13&32.84\\
\hline
\hline
\end{tabular}
	}
\end{table*}

\begin{table}[htbp]
\caption{ The observed yields ($N_{\rm DT}^i$), the produced yields ($N_{\rm prd}^i$) and the determined partial decay rates ($\Delta\Gamma$) of $\kmunu$ in different $q^2$ intervals of data, where the uncertainties are statistical
only. \label{tab:kmunu_decayrate}}
\centering
\resizebox{0.49\textwidth}{!}{
\begin{tabular}{cccc}
\hline
\hline
$q^2~({\rm GeV^2}/c^4)$&$N_{\rm DT}^i$&$N_{\rm prd}^i$&$\Delta\Gamma~({\rm ns^{-1}})$\\ \hline
$(0.01,0.10)$&  $8165\pm104$&$20612\pm274$&$6.341\pm0.084$ \\
$(0.10,0.20)$&  $11686\pm124$&$28015\pm321$&$8.618\pm0.099$ \\
$(0.20,0.30)$&  $11845\pm126$&$26431\pm311$&$8.131\pm0.096$ \\
$(0.30,0.40)$&  $12107\pm130$&$25037\pm304$&$7.702\pm0.093$ \\
$(0.40,0.50)$&  $12238\pm139$&$23450\pm304$&$7.214\pm0.094$ \\
$(0.50,0.60)$&  $12231\pm142$&$21968\pm294$&$6.758\pm0.090$ \\
$(0.60,0.70)$&  $11779\pm135$&$19861\pm265$&$6.110\pm0.081$ \\
$(0.70,0.80)$&  $11329\pm127$&$18313\pm239$&$5.633\pm0.073$ \\
$(0.80,0.90)$&  $10646\pm124$&$16544\pm224$&$5.089\pm0.069$ \\
$(0.90,1.00)$&  $9895\pm122$&$15090\pm216$&$4.642\pm0.066$ \\
$(1.00,1.10)$&  $8619\pm115$&$12877\pm200$&$3.961\pm0.061$ \\
$(1.10,1.20)$&  $7495\pm110$&$11213\pm191$&$3.449\pm0.059$ \\
$(1.20,1.30)$&  $6177\pm106$&$9438\pm188$&$2.903\pm0.058$ \\
$(1.30,1.40)$&  $5077\pm103$&$8035\pm185$&$2.472\pm0.057$ \\
$(1.40,1.50)$&  $3692\pm101$&$5962\pm186$&$1.834\pm0.057$ \\
$(1.50,1.60)$&  $2420\pm67$&$4107\pm129$&$1.263\pm0.040$ \\
$(1.60,1.70)$&  $1475\pm55$&$2719\pm113$&$0.837\pm0.035$ \\
$(1.70,1.88)$&  $706\pm46$&$1771\pm124$&$0.545\pm0.038$ \\
\hline
\hline
\end{tabular}
	}
\end{table}

\begin{table}[htbp]
\caption{ The observed yields ($N_{\rm DT}^i$), the produced yields ($N_{\rm prd}^i$) and the determined partial decay rates ($\Delta\Gamma$) of $\ksenu$ in different $q^2$ intervals of data, where the uncertainties are statistical
only. \label{tab:ksenu_decayrate}}
\centering
\resizebox{0.49\textwidth}{!}{
\begin{tabular}{cccc}
\hline
\hline
$q^2~({\rm GeV^2}/c^4)$&$N_{\rm DT}^i$&$N_{\rm prd}^i$&$\Delta\Gamma~({\rm ns^{-1}})$\\ \hline
$(0.00,0.10)$&  $7031\pm90$&$39669\pm536$&$9.286\pm0.126$ \\
$(0.10,0.20)$&  $6296\pm85$&$36805\pm559$&$8.616\pm0.131$ \\
$(0.20,0.30)$&  $5823\pm81$&$34781\pm556$&$8.142\pm0.130$ \\
$(0.30,0.40)$&  $5425\pm78$&$33376\pm557$&$7.813\pm0.130$ \\
$(0.40,0.50)$&  $4823\pm74$&$29974\pm539$&$7.017\pm0.126$ \\
$(0.50,0.60)$&  $4502\pm71$&$28876\pm531$&$6.760\pm0.124$ \\
$(0.60,0.70)$&  $4033\pm67$&$25941\pm511$&$6.072\pm0.120$ \\
$(0.70,0.80)$&  $3750\pm64$&$24813\pm500$&$5.808\pm0.117$ \\
$(0.80,0.90)$&  $3255\pm60$&$21458\pm467$&$5.023\pm0.109$ \\
$(0.90,1.00)$&  $2962\pm57$&$20048\pm451$&$4.693\pm0.105$ \\
$(1.00,1.10)$&  $2517\pm52$&$17291\pm421$&$4.048\pm0.098$ \\
$(1.10,1.20)$&  $2123\pm48$&$14805\pm390$&$3.466\pm0.091$ \\
$(1.20,1.30)$&  $1756\pm44$&$12433\pm360$&$2.910\pm0.084$ \\
$(1.30,1.40)$&  $1359\pm39$&$9787\pm318$&$2.291\pm0.074$ \\
$(1.40,1.50)$&  $986\pm33$&$7204\pm276$&$1.686\pm0.065$ \\
$(1.50,1.60)$&  $684\pm28$&$5118\pm232$&$1.198\pm0.054$ \\
$(1.60,1.70)$&  $437\pm22$&$3359\pm187$&$0.786\pm0.044$ \\
$(1.70,1.88)$&  $190\pm15$&$1523\pm132$&$0.357\pm0.031$ \\
\hline
\hline
\end{tabular}
	}
\end{table}

\begin{table}[htbp]
\caption{ The observed yields ($N_{\rm DT}^i$), the produced yields ($N_{\rm prd}^i$) and the determined partial decay rates ($\Delta\Gamma$) of $\ksmunu$ in different $q^2$ intervals of data, where the uncertainties are statistical
only. \label{tab:ksmunu_decayrate}}
\centering
\resizebox{0.49\textwidth}{!}{
\begin{tabular}{cccc}
\hline
\hline
$q^2~({\rm GeV^2}/c^4)$&$N_{\rm DT}^i$&$N_{\rm prd}^i$&$\Delta\Gamma~({\rm ns^{-1}})$\\ \hline
$(0.01,0.10)$&  $3044\pm59$&$27122\pm547$&$6.349\pm0.128$ \\
$(0.10,0.20)$&  $4105\pm68$&$36110\pm642$&$8.453\pm0.150$ \\
$(0.20,0.30)$&  $4133\pm69$&$35061\pm640$&$8.208\pm0.150$ \\
$(0.30,0.40)$&  $4066\pm69$&$32504\pm612$&$7.609\pm0.143$ \\
$(0.40,0.50)$&  $4085\pm71$&$31264\pm608$&$7.319\pm0.142$ \\
$(0.50,0.60)$&  $4009\pm72$&$29618\pm603$&$6.933\pm0.141$ \\
$(0.60,0.70)$&  $3588\pm68$&$25397\pm550$&$5.945\pm0.129$ \\
$(0.70,0.80)$&  $3479\pm67$&$24177\pm530$&$5.660\pm0.124$ \\
$(0.80,0.90)$&  $3239\pm66$&$22154\pm520$&$5.186\pm0.122$ \\
$(0.90,1.00)$&  $2911\pm62$&$20078\pm486$&$4.700\pm0.114$ \\
$(1.00,1.10)$&  $2561\pm59$&$17775\pm469$&$4.161\pm0.110$ \\
$(1.10,1.20)$&  $2070\pm57$&$14579\pm462$&$3.413\pm0.108$ \\
$(1.20,1.30)$&  $1762\pm54$&$12513\pm433$&$2.929\pm0.101$ \\
$(1.30,1.40)$&  $1469\pm53$&$10586\pm428$&$2.478\pm0.100$ \\
$(1.40,1.50)$&  $1115\pm46$&$8172\pm383$&$1.913\pm0.090$ \\
$(1.50,1.60)$&  $786\pm38$&$5881\pm318$&$1.377\pm0.075$ \\
$(1.60,1.70)$&  $450\pm28$&$3380\pm233$&$0.791\pm0.055$ \\
$(1.70,1.88)$&  $263\pm21$&$2136\pm189$&$0.500\pm0.044$ \\
\hline
\hline
\end{tabular}
	}
\end{table}

\begin{table*}[htbp]
\caption{Statistical covariance density matrix for $\kenu$. The $i$ denote the reconstructed bin, and
the $j$ represent the produced bin.
\label{tab:kenu_statmatrix}}
\centering
\resizebox{1.0\textwidth}{!}{
\begin{tabular}{c|cccccccccccccccccc}
\hline
\hline
$\rho_{ij}^{\rm stat}$&1&2&3&4&5&6&7&8&9&10&11&12&13&14&15&16&17&18\\ \hline
1&1.000&-0.099&0.004&-0.002&0.000&0.000&0.000&0.000&0.000&0.000&0.000&0.000&0.000&0.000&0.000&0.000&0.000&0.000\\
2&-0.099&1.000&-0.132&0.007&-0.002&0.000&0.000&0.000&0.000&0.000&0.000&0.000&0.000&0.000&0.000&0.000&0.000&0.000\\
3&0.004&-0.132&1.000&-0.148&0.009&-0.003&0.000&0.000&0.000&0.000&0.000&0.000&0.000&0.000&0.000&0.000&0.000&0.000\\
4&-0.002&0.007&-0.148&1.000&-0.158&0.010&-0.002&0.000&0.000&0.000&0.000&0.000&0.000&0.000&0.000&0.000&0.000&0.000\\
5&0.000&-0.002&0.009&-0.158&1.000&-0.162&0.011&-0.002&0.000&0.000&0.000&0.000&0.000&0.000&0.000&0.000&0.000&0.000\\
6&0.000&0.000&-0.003&0.010&-0.162&1.000&-0.165&0.011&-0.003&-0.001&0.000&0.000&0.000&0.000&0.000&0.000&0.000&0.000\\
7&0.000&0.000&0.000&-0.002&0.011&-0.165&1.000&-0.167&0.011&-0.003&0.000&0.000&0.000&0.000&0.000&0.000&0.000&0.000\\
\hline
\hline
\end{tabular}
	}
\end{table*}

\begin{table*}[htbp]
\caption{Statistical covariance density matrix for $\kmunu$. The $i$ denote the reconstructed bin, and
the $j$ represent the produced bin.
\label{tab:kmunu_statmatrix}}
\centering
\resizebox{1.0\textwidth}{!}{
\begin{tabular}{c|cccccccccccccccccc}
\hline
\hline
$\rho_{ij}^{\rm stat}$&1&2&3&4&5&6&7&8&9&10&11&12&13&14&15&16&17&18\\ \hline
1&1.000&-0.068&0.003&0.000&0.000&0.000&0.000&0.000&0.000&0.000&0.000&0.000&0.000&0.000&0.000&0.000&0.000&0.000\\
2&-0.068&1.000&-0.088&0.005&-0.001&0.000&0.000&0.000&0.000&0.000&0.000&0.000&0.000&0.000&0.000&0.000&0.000&0.000\\
3&0.003&-0.088&1.000&-0.105&0.007&-0.001&0.000&0.000&0.000&0.000&0.000&0.000&0.000&0.000&0.000&0.000&0.000&0.000\\
4&0.000&0.005&-0.105&1.000&-0.118&0.008&-0.001&0.000&0.000&0.000&0.000&0.000&0.000&0.000&0.000&0.000&0.000&0.000\\
5&0.000&-0.001&0.007&-0.118&1.000&-0.128&0.008&-0.002&0.000&0.000&0.000&0.000&0.000&0.000&0.000&0.000&0.000&0.000\\
6&0.000&0.000&-0.001&0.008&-0.128&1.000&-0.134&0.008&-0.002&-0.001&0.000&0.000&0.000&0.000&0.000&0.000&0.000&0.000\\
7&0.000&0.000&0.000&-0.001&0.008&-0.134&1.000&-0.138&0.008&-0.002&-0.001&0.000&0.000&0.000&0.000&0.000&0.000&0.000\\
\hline
\hline
\end{tabular}
	}
\end{table*}

\begin{table*}[htbp]
\caption{Statistical covariance density matrix for $\ksenu$. The $i$ denote the reconstructed bin, and
the $j$ represent the produced bin.
\label{tab:ksenu_statmatrix}}
\centering
\resizebox{1.0\textwidth}{!}{
\begin{tabular}{c|cccccccccccccccccc}
\hline
\hline
$\rho_{ij}^{\rm stat}$&1&2&3&4&5&6&7&8&9&10&11&12&13&14&15&16&17&18\\ \hline
1&1.000&-0.089&0.004&-0.001&0.000&0.000&0.000&0.000&0.000&0.000&0.000&0.000&0.000&0.000&0.000&0.000&0.000&0.000\\
2&-0.089&1.000&-0.117&0.006&-0.002&0.000&0.000&0.000&0.000&0.000&0.000&0.000&0.000&0.000&0.000&0.000&0.000&0.000\\
3&0.004&-0.117&1.000&-0.133&0.009&-0.002&0.000&0.000&0.000&0.000&0.000&0.000&0.000&0.000&0.000&0.000&0.000&0.000\\
4&-0.001&0.006&-0.133&1.000&-0.141&0.010&-0.001&0.000&0.000&0.000&0.000&0.000&0.000&0.000&0.000&0.000&0.000&0.000\\
5&0.000&-0.002&0.009&-0.141&1.000&-0.149&0.012&-0.001&0.000&0.000&0.000&0.000&0.000&0.000&0.000&0.000&0.000&0.000\\
6&0.000&0.000&-0.002&0.010&-0.149&1.000&-0.152&0.013&-0.001&0.000&0.000&0.000&0.000&0.000&0.000&0.000&0.000&0.000\\
7&0.000&0.000&0.000&-0.001&0.012&-0.152&1.000&-0.156&0.013&-0.001&0.000&0.000&0.000&0.000&0.000&0.000&0.000&0.000\\
\hline
\hline
\end{tabular}
	}
\end{table*}

\begin{table*}[htbp]
\caption{Statistical covariance density matrix for $\ksmunu$. The $i$ denote the reconstructed bin, and
the $j$ represent the produced bin.
\label{tab:ksmunu_statmatrix}}
\centering
\resizebox{1.0\textwidth}{!}{
\begin{tabular}{c|cccccccccccccccccc}
\hline
\hline
$\rho_{ij}^{\rm stat}$&1&2&3&4&5&6&7&8&9&10&11&12&13&14&15&16&17&18\\ \hline
1&1.000&-0.063&0.003&0.000&0.000&0.000&0.000&0.000&0.000&0.000&0.000&0.000&0.000&0.000&0.000&0.000&0.000&-0.001\\
2&-0.063&1.000&-0.080&0.005&0.000&0.000&0.000&0.000&0.000&0.000&0.000&0.000&0.000&0.000&-0.001&0.000&-0.001&-0.001\\
3&0.003&-0.080&1.000&-0.096&0.007&0.000&0.000&0.000&0.000&0.000&0.000&0.000&0.000&0.000&0.000&0.000&0.000&-0.001\\
4&0.000&0.005&-0.096&1.000&-0.109&0.008&0.000&0.000&0.000&0.000&0.000&0.000&0.000&0.000&0.000&0.000&0.000&-0.001\\
5&0.000&0.000&0.007&-0.109&1.000&-0.116&0.009&-0.001&0.000&0.000&0.000&0.000&0.000&0.000&0.000&-0.001&-0.001&-0.001\\
6&0.000&0.000&0.000&0.008&-0.116&1.000&-0.121&0.010&-0.001&0.000&0.000&0.000&0.000&0.000&0.000&0.000&-0.001&-0.001\\
7&0.000&0.000&0.000&0.000&0.009&-0.121&1.000&-0.126&0.010&-0.001&0.000&0.000&0.000&0.000&-0.001&0.000&-0.001&0.000\\
\hline
\hline
\end{tabular}
	}
\end{table*}

\begin{table*}[htbp]
\caption{The systematic uncertainties (in \%) of the measured decay rates of $\kmunu$ in different
$q^2$ bins.
\label{tab:kmunu_sysq2}}
\centering
\resizebox{1.0\textwidth}{!}{
\begin{tabular}{ccccccccccccccccccc}
\hline
\hline
$i$-th $q^2$ bin&1&2&3&4&5&6&7&8&9&10&11&12&13&14&15&16&17&18\\ \hline
$N_{\rm ST}$&0.30&0.30&0.30&0.30&0.30&0.30&0.30&0.30&0.30&0.30&0.30&0.30&0.30&0.30&0.30&0.30&0.30&0.30\\
$D^0$ lifetime&0.24&0.24&0.24&0.24&0.24&0.24&0.24&0.24&0.24&0.24&0.24&0.24&0.24&0.24&0.24&0.24&0.24&0.24\\
MC statistics&0.25&0.21&0.20&0.20&0.20&0.20&0.20&0.20&0.21&0.21&0.23&0.24&0.27&0.31&0.36&0.44&0.58&0.93\\
MC model&0.18&0.21&0.71&0.44&0.20&0.23&0.28&0.06&0.19&0.20&0.16&0.74&1.03&0.87&0.21&0.74&1.02&1.06\\
$K^-$ tracking&0.10&0.10&0.10&0.10&0.10&0.10&0.10&0.10&0.10&0.10&0.10&0.11&0.11&0.13&0.14&0.18&0.23&0.43\\
$K^-$ PID&0.10&0.10&0.10&0.10&0.10&0.10&0.10&0.10&0.10&0.10&0.10&0.10&0.10&0.10&0.10&0.10&0.10&0.18\\
$\mu^+$ tracking&0.10&0.10&0.10&0.10&0.10&0.10&0.10&0.10&0.10&0.10&0.10&0.10&0.10&0.10&0.10&0.10&0.10&0.10\\
$\mu^+$ PID&0.20&0.20&0.19&0.18&0.18&0.17&0.17&0.17&0.16&0.17&0.17&0.18&0.18&0.19&0.19&0.20&0.22&0.23\\
$U_{\rm miss}$ fit&0.24&0.11&0.11&0.11&0.10&0.14&0.12&0.13&0.15&0.15&0.11&0.15&0.11&0.18&0.17&0.21&0.18&0.26\\
$E_{\rm extra~\gamma}^{\rm max}$ and $N_{\rm extra}^{\rm trk}$ cut&0.10&0.10&0.10&0.10&0.10&0.10&0.10&0.10&0.10&0.10&0.10&0.10&0.10&0.10&0.10&0.10&0.10&0.10\\

\hline
Total&0.62&0.57&0.88&0.68&0.55&0.57&0.59&0.53&0.56&0.57&0.56&0.92&1.17&1.06&0.66&1.02&1.30&1.58\\
\hline
\hline
\end{tabular}
	}
\end{table*}

\begin{table*}[htbp]
\caption{The systematic uncertainties (in \%) of the measured decay rates of $\ksenu$ in different
$q^2$ bins.
\label{tab:ksenu_sysq2}}
\centering
\resizebox{1.0\textwidth}{!}{
\begin{tabular}{ccccccccccccccccccc}
\hline
\hline
$i$-th $q^2$ bin&1&2&3&4&5&6&7&8&9&10&11&12&13&14&15&16&17&18\\ \hline
$N_{\rm tag}$&0.30&0.30&0.30&0.30&0.30&0.30&0.30&0.30&0.30&0.30&0.30&0.30&0.30&0.30&0.30&0.30&0.30&0.30\\
$D^+$ lifetime&0.48&0.48&0.48&0.48&0.48&0.48&0.48&0.48&0.48&0.48&0.48&0.48&0.48&0.48&0.48&0.48&0.48&0.48\\
MC statistics&0.19&0.19&0.20&0.21&0.22&0.23&0.25&0.26&0.27&0.29&0.32&0.34&0.38&0.43&0.49&0.59&0.77&1.09\\
MC model&0.32&0.66&0.70&0.22&0.91&0.60&1.11&0.27&1.14&0.38&0.90&0.01&0.99&0.02&0.39&2.18&0.17&0.46\\
$e^+$ tracking&0.10&0.10&0.10&0.10&0.10&0.10&0.10&0.10&0.10&0.10&0.10&0.10&0.10&0.10&0.10&0.10&0.10&0.10\\
$e^+$ PID&0.10&0.10&0.10&0.10&0.10&0.10&0.10&0.10&0.10&0.10&0.10&0.10&0.10&0.10&0.10&0.10&0.10&0.10\\
$U_{\rm miss}$ fit&0.53&1.34&0.26&0.55&1.65&1.29&0.04&0.58&0.20&0.05&0.22&0.17&0.84&0.05&0.04&0.05&0.32&0.06\\
$E_{\rm extra~\gamma}^{\rm max}$ and $N_{\rm extra}^{\rm trk}$ cut&0.10&0.10&0.10&0.10&0.10&0.10&0.10&0.10&0.10&0.10&0.10&0.10&0.10&0.10&0.10&0.10&0.10&0.10\\
$K_S^0$ reconstruction&0.75&0.72&0.68&0.65&0.60&0.60&0.70&0.82&0.95&1.11&1.31&1.44&1.48&1.55&1.62&1.72&1.75&1.93\\
Quoted branching fractions &0.07&0.07&0.07&0.07&0.07&0.07&0.07&0.07&0.07&0.07&0.07&0.07&0.07&0.07&0.07&0.07&0.07&0.07\\

\hline
Total&1.16&1.77&1.19&1.08&2.08&1.67&1.46&1.23&1.64&1.35&1.74&1.60&2.09&1.71&1.84&2.90&2.03&2.34\\
\hline
\hline
\end{tabular}
	}
\end{table*}

\begin{table*}[htbp]
\caption{The systematic uncertainties (in \%) of the measured decay rates of $\ksmunu$ in different
$q^2$ bins.
\label{tab:ksmunu_sysq2}}
\centering
\resizebox{1.0\textwidth}{!}{
\begin{tabular}{ccccccccccccccccccc}
\hline
\hline
$i$-th $q^2$ bin&1&2&3&4&5&6&7&8&9&10&11&12&13&14&15&16&17&18\\ \hline
$N_{\rm tag}$&0.30&0.30&0.30&0.30&0.30&0.30&0.30&0.30&0.30&0.30&0.30&0.30&0.30&0.30&0.30&0.30&0.30&0.30\\
$D^+$ lifetime&0.48&0.48&0.48&0.48&0.48&0.48&0.48&0.48&0.48&0.48&0.48&0.48&0.48&0.48&0.48&0.48&0.48&0.48\\
MC statistics&0.29&0.25&0.25&0.25&0.25&0.25&0.26&0.26&0.28&0.29&0.31&0.34&0.38&0.42&0.48&0.58&0.74&1.02\\
MC model&0.71&0.18&0.32&0.79&0.28&0.18&0.61&0.26&0.76&0.33&1.53&0.77&0.78&1.41&0.74&0.23&1.47&1.75\\
$\mu^+$ tracking&0.10&0.10&0.10&0.10&0.10&0.10&0.10&0.10&0.10&0.10&0.10&0.10&0.10&0.10&0.10&0.10&0.10&0.10\\
$\mu^+$ PID&0.17&0.17&0.17&0.17&0.16&0.16&0.16&0.16&0.16&0.16&0.17&0.17&0.18&0.19&0.18&0.19&0.22&0.25\\
$U_{\rm miss}$ fit&0.03&0.10&0.03&0.23&0.25&0.05&0.19&0.17&0.07&0.07&0.06&0.06&0.07&0.05&0.06&0.13&0.10&0.18\\
$E_{\rm extra~\gamma}^{\rm max}$ and $N_{\rm extra}^{\rm trk}$ cut&0.10&0.10&0.10&0.10&0.10&0.10&0.10&0.10&0.10&0.10&0.10&0.10&0.10&0.10&0.10&0.10&0.10&0.10\\
$K_S^0$ reconstruction&1.05&0.71&0.68&0.65&0.60&0.60&0.70&0.82&0.95&1.11&1.30&1.44&1.48&1.55&1.62&1.72&1.77&1.96\\
Quoted branching fractions&0.07&0.07&0.07&0.07&0.07&0.07&0.07&0.07&0.07&0.07&0.07&0.07&0.07&0.07&0.07&0.07&0.07&0.07\\

\hline
Total&1.44&0.99&1.00&1.24&0.97&0.91&1.16&1.10&1.39&1.34&2.12&1.78&1.82&2.22&1.95&1.93&2.50&2.90\\
\hline
\hline
\end{tabular}
	}
\end{table*}

\begin{table*}[htbp]
\caption{ Systematic covariance density matrix for $\kenu$, where $i$ denotes the reconstructed bin, and
the $j$ represent the produced bin.
\label{tab:kenu_sysmatrix}}
\centering
\resizebox{1.0\textwidth}{!}{
\begin{tabular}{c|cccccccccccccccccc}
\hline
\hline
$\rho_{ij}^{\rm sys}$&1&2&3&4&5&6&7&8&9&10&11&12&13&14&15&16&17&18\\ \hline
1&1.000&0.773&0.747&0.789&0.730&0.762&0.586&0.767&0.580&0.774&0.759&0.680&0.719&0.600&0.683&0.626&0.474&0.491\\
2&0.773&1.000&0.704&0.782&0.733&0.745&0.556&0.759&0.549&0.763&0.743&0.685&0.700&0.571&0.673&0.618&0.441&0.472\\
3&0.747&0.704&1.000&0.756&0.578&0.842&0.851&0.740&0.849&0.783&0.832&0.518&0.838&0.845&0.686&0.616&0.784&0.634\\
4&0.789&0.782&0.756&1.000&0.658&0.788&0.655&0.764&0.650&0.780&0.781&0.643&0.753&0.664&0.687&0.626&0.554&0.529\\
5&0.730&0.733&0.578&0.658&1.000&0.580&0.351&0.686&0.341&0.670&0.618&0.693&0.555&0.373&0.594&0.552&0.224&0.341\\
6&0.762&0.745&0.842&0.788&0.580&1.000&0.777&0.751&0.787&0.784&0.817&0.559&0.812&0.789&0.688&0.620&0.712&0.601\\
7&0.586&0.556&0.851&0.655&0.351&0.777&1.000&0.588&0.969&0.674&0.773&0.291&0.821&0.944&0.586&0.514&0.951&0.670\\
\hline
\hline
\end{tabular}
	}
\end{table*}

\begin{table*}[htbp]
\caption{ Systematic covariance density matrix for $\kmunu$, where $i$ denotes the reconstructed bin, and
the $j$ represent the produced bin.
\label{tab:kmunu_sysmatrix}}
\centering
\resizebox{1.0\textwidth}{!}{
\begin{tabular}{c|cccccccccccccccccc}
\hline
\hline
$\rho_{ij}^{\rm sys}$&1&2&3&4&5&6&7&8&9&10&11&12&13&14&15&16&17&18\\ \hline
1&1.000&0.717&0.637&0.702&0.731&0.716&0.717&0.679&0.697&0.695&0.697&0.611&0.555&0.573&0.614&0.561&0.512&0.456\\
2&0.717&1.000&0.718&0.790&0.804&0.790&0.796&0.735&0.766&0.765&0.763&0.701&0.646&0.662&0.676&0.643&0.593&0.527\\
3&0.637&0.718&1.000&0.853&0.714&0.729&0.770&0.536&0.673&0.685&0.652&0.894&0.904&0.881&0.606&0.807&0.809&0.703\\
4&0.702&0.790&0.853&1.000&0.759&0.783&0.808&0.644&0.739&0.745&0.725&0.838&0.820&0.813&0.658&0.761&0.740&0.648\\
5&0.731&0.804&0.714&0.759&1.000&0.759&0.792&0.737&0.764&0.762&0.762&0.686&0.629&0.646&0.673&0.629&0.578&0.514\\
6&0.716&0.790&0.729&0.783&0.759&1.000&0.754&0.714&0.749&0.749&0.745&0.702&0.652&0.665&0.661&0.643&0.597&0.530\\
7&0.717&0.796&0.770&0.808&0.792&0.754&1.000&0.668&0.754&0.753&0.746&0.743&0.701&0.709&0.665&0.678&0.638&0.565\\
\hline
\hline
\end{tabular}
	}
\end{table*}

\begin{table*}[htbp]
\caption{ Systematic covariance density matrix for $\ksenu$, where $i$ denotes the reconstructed bin, and
the $j$ represent the produced bin.
\label{tab:ksenu_sysmatrix}}
\centering
\resizebox{1.0\textwidth}{!}{
\begin{tabular}{c|cccccccccccccccccc}
\hline
\hline
$\rho_{ij}^{\rm sys}$&1&2&3&4&5&6&7&8&9&10&11&12&13&14&15&16&17&18\\ \hline
1&1.000&0.629&0.686&0.645&0.573&0.593&0.620&0.632&0.647&0.662&0.677&0.613&0.651&0.579&0.586&0.572&0.464&0.399\\
2&0.629&1.000&0.650&0.573&0.604&0.586&0.666&0.559&0.674&0.591&0.655&0.494&0.630&0.468&0.511&0.617&0.386&0.348\\
3&0.686&0.650&1.000&0.547&0.647&0.618&0.711&0.580&0.715&0.611&0.686&0.500&0.659&0.474&0.525&0.655&0.393&0.358\\
4&0.645&0.573&0.547&1.000&0.428&0.523&0.525&0.562&0.547&0.580&0.578&0.544&0.554&0.513&0.510&0.472&0.407&0.346\\
5&0.573&0.604&0.647&0.428&1.000&0.484&0.659&0.477&0.647&0.504&0.595&0.379&0.571&0.360&0.424&0.601&0.305&0.288\\
6&0.593&0.586&0.618&0.523&0.484&1.000&0.549&0.507&0.603&0.524&0.581&0.432&0.556&0.408&0.448&0.546&0.336&0.304\\
7&0.620&0.666&0.711&0.525&0.659&0.549&1.000&0.461&0.728&0.550&0.663&0.408&0.637&0.388&0.466&0.682&0.333&0.318\\
\hline
\hline
\end{tabular}
	}
\end{table*}

\begin{table*}[htbp]
\caption{ Systematic covariance density matrix for $\ksmunu$, where $i$ denotes the reconstructed bin, and
the $j$ represent the produced bin.
\label{tab:ksmunu_sysmatrix}}
\centering
\resizebox{1.0\textwidth}{!}{
\begin{tabular}{c|cccccccccccccccccc}
\hline
\hline
$\rho_{ij}^{\rm sys}$&1&2&3&4&5&6&7&8&9&10&11&12&13&14&15&16&17&18\\ \hline
1&1.000&0.587&0.633&0.674&0.590&0.560&0.656&0.620&0.699&0.653&0.729&0.705&0.691&0.703&0.640&0.539&0.583&0.532\\
2&0.587&1.000&0.557&0.546&0.561&0.554&0.557&0.585&0.578&0.599&0.539&0.591&0.579&0.538&0.541&0.501&0.448&0.401\\
3&0.633&0.557&1.000&0.553&0.569&0.550&0.586&0.583&0.608&0.597&0.592&0.608&0.595&0.578&0.552&0.487&0.478&0.431\\
4&0.674&0.546&0.553&1.000&0.511&0.516&0.638&0.554&0.664&0.568&0.713&0.632&0.617&0.665&0.561&0.435&0.543&0.499\\
5&0.590&0.561&0.569&0.511&1.000&0.472&0.556&0.551&0.568&0.559&0.546&0.564&0.551&0.532&0.511&0.452&0.439&0.395\\
6&0.560&0.554&0.550&0.516&0.472&1.000&0.470&0.545&0.535&0.548&0.498&0.539&0.527&0.493&0.491&0.450&0.408&0.365\\
7&0.656&0.557&0.586&0.638&0.556&0.470&1.000&0.511&0.643&0.577&0.665&0.622&0.608&0.631&0.558&0.456&0.518&0.474\\
\hline
\hline
\end{tabular}
	}
\end{table*}

\begin{sidewaystable}[htbp]
\caption{The elements of the covariance density matrix $\rho_{ij}(i=0,1,2...,35,36; j=0,1,2...,35,36;)$ for the simultaneous fit.}
\label{klnumatrix_cov1}
\centering
\resizebox{1.0\textwidth}{!}{
\begin{tabular}{c|cccccccccccccccccccccccccccccccccccc}
\hline
\hline
$\rho_{ij}$&1&2&3&4&5&6&7&8&9&10&11&12&13&14&15&16&17&18&19&20&21&22&23&24&25&26&27&28&29&30&31&32&33&34&35&36\\ \hline
1&1.000&0.085&0.217&0.166&0.122&0.181&0.241&0.138&0.236&0.138&0.147&0.084&0.142&0.184&0.087&0.076&0.192&0.070&0.018&0.016&0.016&0.016&0.014&0.014&0.013&0.013&0.011&0.012&0.010&0.010&0.009&0.008&0.007&0.006&0.005&0.003\\
2&0.085&1.000&0.087&0.157&0.110&0.161&0.208&0.124&0.203&0.124&0.131&0.078&0.125&0.159&0.078&0.068&0.164&0.062&0.017&0.015&0.015&0.015&0.013&0.013&0.012&0.012&0.011&0.011&0.010&0.009&0.008&0.008&0.007&0.005&0.005&0.003\\
3&0.217&0.087&1.000&0.095&0.120&0.250&0.424&0.166&0.417&0.175&0.203&0.073&0.208&0.317&0.110&0.093&0.369&0.111&0.014&0.012&0.012&0.013&0.011&0.011&0.010&0.011&0.009&0.009&0.008&0.008&0.007&0.007&0.006&0.005&0.004&0.003\\
4&0.166&0.157&0.095&1.000&-0.033&0.182&0.249&0.127&0.245&0.129&0.141&0.072&0.138&0.189&0.081&0.070&0.206&0.070&0.015&0.013&0.013&0.014&0.011&0.012&0.011&0.011&0.010&0.010&0.009&0.008&0.007&0.007&0.006&0.005&0.004&0.003\\
5&0.122&0.110&0.120&-0.033&1.000&-0.037&0.102&0.087&0.092&0.084&0.083&0.066&0.074&0.076&0.054&0.048&0.063&0.034&0.015&0.013&0.013&0.013&0.011&0.012&0.011&0.011&0.010&0.010&0.009&0.008&0.007&0.007&0.006&0.005&0.004&0.003\\
6&0.181&0.161&0.250&0.182&-0.037&1.000&0.244&0.150&0.325&0.145&0.166&0.067&0.168&0.249&0.092&0.078&0.285&0.088&0.014&0.012&0.012&0.012&0.010&0.011&0.009&0.010&0.009&0.009&0.008&0.007&0.007&0.006&0.005&0.004&0.004&0.003\\
7&0.241&0.208&0.424&0.249&0.102&0.244&1.000&0.100&0.634&0.214&0.266&0.056&0.283&0.472&0.134&0.110&0.577&0.157&0.009&0.008&0.008&0.008&0.007&0.007&0.006&0.006&0.006&0.006&0.005&0.005&0.004&0.004&0.004&0.003&0.002&0.002\\
\hline
\hline
\end{tabular}

	}
\end{sidewaystable}

\begin{sidewaystable}[htbp]
\caption{The elements of the covariance density matrix $\rho_{ij}(i=0,1,2...,35,36; j=37,38,39...,71,72;)$ for the simultaneous fit.}
\label{klnumatrix_cov2}
\centering
\resizebox{1.0\textwidth}{!}{
\begin{tabular}{c|cccccccccccccccccccccccccccccccccccc}
\hline
\hline
$\rho_{ij}$&37&38&39&40&41&42&43&44&45&46&47&48&49&50&51&52&53&54&55&56&57&58&59&60&61&62&63&64&65&66&67&68&69&70&71&72\\ \hline
1&0.069&0.064&0.054&0.058&0.057&0.051&0.033&0.050&0.031&0.046&0.043&0.042&0.037&0.029&0.030&0.027&0.015&0.016&0.004&0.003&0.003&0.003&0.003&0.003&0.003&0.003&0.002&0.003&0.002&0.002&0.002&0.002&0.002&0.001&0.001&0.001\\
2&0.079&0.074&0.062&0.067&0.066&0.059&0.038&0.057&0.036&0.053&0.049&0.048&0.042&0.033&0.034&0.032&0.017&0.019&0.005&0.004&0.004&0.004&0.003&0.003&0.003&0.003&0.003&0.003&0.003&0.002&0.002&0.002&0.002&0.001&0.001&0.001\\
3&0.068&0.064&0.054&0.058&0.057&0.051&0.033&0.050&0.031&0.046&0.043&0.042&0.037&0.029&0.030&0.027&0.015&0.016&0.004&0.003&0.003&0.003&0.003&0.003&0.003&0.003&0.002&0.003&0.002&0.002&0.002&0.002&0.002&0.001&0.001&0.001\\
4&0.072&0.068&0.057&0.061&0.061&0.054&0.035&0.053&0.033&0.049&0.045&0.044&0.039&0.031&0.032&0.029&0.016&0.017&0.004&0.004&0.004&0.004&0.003&0.003&0.003&0.003&0.003&0.003&0.002&0.002&0.002&0.002&0.002&0.001&0.001&0.001\\
5&0.071&0.067&0.056&0.060&0.060&0.053&0.035&0.052&0.033&0.048&0.044&0.044&0.038&0.030&0.031&0.029&0.015&0.017&0.004&0.004&0.003&0.004&0.003&0.003&0.003&0.003&0.003&0.003&0.002&0.002&0.002&0.002&0.002&0.001&0.001&0.001\\
6&0.069&0.065&0.054&0.058&0.058&0.052&0.033&0.050&0.032&0.046&0.043&0.042&0.037&0.029&0.030&0.028&0.015&0.016&0.004&0.003&0.003&0.003&0.003&0.003&0.003&0.003&0.002&0.003&0.002&0.002&0.002&0.002&0.002&0.001&0.001&0.001\\
7&0.069&0.064&0.054&0.058&0.058&0.051&0.033&0.050&0.032&0.046&0.043&0.042&0.037&0.029&0.030&0.028&0.015&0.016&0.004&0.003&0.003&0.003&0.003&0.003&0.003&0.003&0.002&0.003&0.002&0.002&0.002&0.002&0.002&0.001&0.001&0.001\\
\hline
\hline
\end{tabular}
	}
\end{sidewaystable}

\begin{sidewaystable}[htbp]
\caption{The elements of the covariance density matrix $\rho_{ij}(i=37,38,39...,71,72; j=0,1,2...,35,36;)$ for the simultaneous fit.}
\label{klnumatrix_cov3}
\centering
\resizebox{1.0\textwidth}{!}{
\begin{tabular}{c|cccccccccccccccccccccccccccccccccccc}
\hline
\hline
$\rho_{ij}$&1&2&3&4&5&6&7&8&9&10&11&12&13&14&15&16&17&18&19&20&21&22&23&24&25&26&27&28&29&30&31&32&33&34&35&36\\ \hline
37&0.069&0.079&0.068&0.072&0.071&0.069&0.069&0.071&0.068&0.065&0.061&0.052&0.044&0.040&0.032&0.033&0.026&0.020&0.004&0.005&0.005&0.005&0.005&0.005&0.004&0.004&0.004&0.004&0.003&0.003&0.003&0.002&0.002&0.002&0.001&0.001\\
38&0.064&0.074&0.064&0.068&0.067&0.065&0.064&0.067&0.064&0.061&0.057&0.049&0.041&0.038&0.030&0.030&0.024&0.019&0.004&0.005&0.005&0.004&0.004&0.004&0.004&0.004&0.004&0.003&0.003&0.003&0.002&0.002&0.002&0.002&0.001&0.001\\
39&0.054&0.062&0.054&0.057&0.056&0.054&0.054&0.056&0.054&0.051&0.048&0.041&0.034&0.032&0.026&0.026&0.020&0.016&0.003&0.004&0.004&0.004&0.004&0.004&0.003&0.003&0.003&0.003&0.002&0.002&0.002&0.002&0.002&0.001&0.001&0.001\\
40&0.058&0.067&0.058&0.061&0.060&0.058&0.058&0.060&0.058&0.055&0.051&0.044&0.037&0.034&0.027&0.027&0.022&0.017&0.004&0.004&0.004&0.004&0.004&0.004&0.004&0.004&0.003&0.003&0.003&0.002&0.002&0.002&0.002&0.002&0.001&0.001\\
41&0.057&0.066&0.057&0.061&0.060&0.058&0.058&0.060&0.057&0.054&0.051&0.044&0.036&0.034&0.027&0.027&0.022&0.017&0.003&0.004&0.004&0.004&0.004&0.004&0.004&0.004&0.003&0.003&0.003&0.002&0.002&0.002&0.002&0.002&0.001&0.001\\
42&0.051&0.059&0.051&0.054&0.053&0.052&0.051&0.053&0.051&0.049&0.045&0.039&0.033&0.030&0.024&0.024&0.019&0.015&0.003&0.004&0.004&0.003&0.004&0.003&0.003&0.003&0.003&0.003&0.002&0.002&0.002&0.002&0.002&0.001&0.001&0.001\\
43&0.033&0.038&0.033&0.035&0.035&0.033&0.033&0.035&0.033&0.032&0.029&0.025&0.021&0.020&0.016&0.016&0.013&0.010&0.002&0.002&0.002&0.002&0.002&0.002&0.002&0.002&0.002&0.002&0.001&0.001&0.001&0.001&0.001&0.001&0.001&0.001\\
\hline
\hline
\end{tabular}
	}
\end{sidewaystable}

\begin{sidewaystable}[htbp]
\caption{The elements of the covariance density matrix $\rho_{ij}(i=37,38,39...,71,72; j=37,38,39...,71,72;)$ for the simultaneous fit.}
\label{klnumatrix_cov4}
\centering
\resizebox{1.0\textwidth}{!}{
\begin{tabular}{c|cccccccccccccccccccccccccccccccccccc}
\hline
\hline
$\rho_{ij}$&37&38&39&40&41&42&43&44&45&46&47&48&49&50&51&52&53&54&55&56&57&58&59&60&61&62&63&64&65&66&67&68&69&70&71&72\\ \hline
37&1.000&0.038&0.129&0.109&0.085&0.085&0.088&0.072&0.080&0.077&0.069&0.096&0.096&0.080&0.039&0.058&0.053&0.035&0.005&0.007&0.006&0.006&0.006&0.006&0.005&0.005&0.005&0.005&0.004&0.004&0.003&0.003&0.003&0.002&0.002&0.001\\
38&0.038&1.000&0.091&0.139&0.102&0.102&0.107&0.084&0.096&0.093&0.083&0.121&0.123&0.102&0.047&0.072&0.067&0.045&0.006&0.007&0.007&0.007&0.007&0.007&0.006&0.006&0.006&0.006&0.005&0.004&0.004&0.003&0.003&0.003&0.002&0.002\\
39&0.129&0.091&1.000&0.145&0.141&0.141&0.157&0.086&0.125&0.124&0.105&0.235&0.256&0.204&0.062&0.137&0.136&0.088&0.005&0.006&0.006&0.006&0.006&0.006&0.005&0.005&0.005&0.005&0.004&0.004&0.003&0.003&0.003&0.002&0.002&0.001\\
40&0.109&0.139&0.145&1.000&0.013&0.125&0.127&0.082&0.107&0.105&0.091&0.173&0.184&0.148&0.053&0.101&0.099&0.064&0.006&0.007&0.007&0.006&0.006&0.006&0.006&0.006&0.005&0.005&0.004&0.004&0.004&0.003&0.003&0.002&0.002&0.002\\
41&0.085&0.102&0.141&0.013&1.000&-0.028&0.100&0.072&0.082&0.080&0.072&0.103&0.104&0.086&0.040&0.061&0.057&0.038&0.006&0.007&0.007&0.006&0.006&0.006&0.006&0.006&0.005&0.005&0.004&0.004&0.004&0.003&0.003&0.002&0.002&0.001\\
42&0.085&0.102&0.141&0.125&-0.028&1.000&-0.028&0.079&0.081&0.079&0.071&0.108&0.110&0.091&0.040&0.064&0.060&0.040&0.005&0.007&0.006&0.006&0.006&0.006&0.005&0.005&0.005&0.005&0.004&0.004&0.003&0.003&0.003&0.002&0.002&0.001\\
43&0.088&0.107&0.157&0.127&0.100&-0.028&1.000&-0.054&0.093&0.082&0.074&0.120&0.124&0.102&0.042&0.071&0.067&0.044&0.005&0.007&0.006&0.006&0.006&0.006&0.005&0.005&0.005&0.005&0.004&0.004&0.003&0.003&0.003&0.002&0.002&0.001\\
\hline
\hline
\end{tabular}
	}
\end{sidewaystable}


\begin{thebibliography}{}


\bibitem{Lubicz:2017syv}
V.~Lubicz \textit{et al.},
\href{https://journals.aps.org/prd/abstract/10.1103/PhysRevD.96.054514}{Phys. Rev. D \textbf{96}, 054514 (2017).}


\bibitem{Chakraborty:2021qav}
B.~Chakraborty \textit{et al.} (HPQCD Collaboration),
\href{https://journals.aps.org/prd/abstract/10.1103/PhysRevD.104.034505}{Phys. Rev. D \textbf{104}, 034505 (2021).}


\bibitem{Parrott:2022rgu}
W.~G.~Parrott \textit{et al.} (HPQCD collaboration),
\href{https://journals.aps.org/prd/abstract/10.1103/PhysRevD.107.014510}{Phys. Rev. D \textbf{107}, 014510 (2023).}


\bibitem{FermilabLattice:2022gku}
A.~Bazavov \textit{et al.} (Fermilab Lattice and MILC Collaborations),
\href{https://journals.aps.org/prd/abstract/10.1103/PhysRevD.107.094516}{Phys. Rev. D \textbf{107}, 094516 (2023).}


\bibitem{Wu:2006rd}
Y.~L.~Wu, M.~Zhong and Y.~B.~Zuo,
\href{https://www.worldscientific.com/doi/abs/10.1142/S0217751X06033209}{Int. J. Mod. Phys. A \textbf{21}, 6125-6172 (2006).}


\bibitem{Verma:2011yw}
R.~C.~Verma,
\href{https://iopscience.iop.org/article/10.1088/0954-3899/39/2/025005}{J. Phys. G \textbf{39}, 025005 (2012).}


\bibitem{Ivanov:2019nqd}
M.~A.~Ivanov, J.~G.~K\"orner, J.~N.~Pandya, P.~Santorelli, N.~R.~Soni and C.~T.~Tran,
\href{https://link.springer.com/article/10.1007/s11467-019-0908-1}{Front. Phys. (Beijing) \textbf{14}, 64401 (2019).}


\bibitem{Faustov:2019mqr}
R.~N.~Faustov, V.~O.~Galkin and X.~W.~Kang,
\href{https://journals.aps.org/prd/abstract/10.1103/PhysRevD.101.013004}{Phys. Rev. D \textbf{101}, 013004 (2020).}


\bibitem{Ke:2023qzc}
B.~C.~Ke, J.~Koponen, H.~B.~Li and Y.~Zheng,
\href{https://www.annualreviews.org/doi/10.1146/annurev-nucl-110222-044046}{Ann. Rev. Nucl. Part. Sci. \textbf{73}, 285-314 (2023).}

\bibitem{pdg2022}
R. L. Workman {\it et al.} (Particle Data Group),
\href{https://pdglive.lbl.gov/Viewer.action}{Prog. Theor. Exp. Phys. {\bf 2022}, 083C01 (2022).}


\bibitem{BES:2004rav}
M.~Ablikim \textit{et al.} (BES Collaboration),
\href{https://www.sciencedirect.com/science/article/pii/S0370269304010123?via\%3Dihub}{Phys. Lett. B \textbf{597}, 39-46 (2004).}


\bibitem{BES:2004obp}
M.~Ablikim \textit{et al.} (BES Collaboration),
\href{https://www.sciencedirect.com/science/article/pii/S0370269304017253?via\%3Dihub}{Phys. Lett. B \textbf{608}, 24-30 (2005).}



\bibitem{BES:2006kzp}
M.~Ablikim \textit{et al.} (BES Collaboration),
\href{https://www.sciencedirect.com/science/article/abs/pii/S0370269306014213?via\%3Dihub}{Phys. Lett. B \textbf{644}, 20-24 (2007).}


\bibitem{BaBar:2007zgf}
B.~Aubert \textit{et al.} (BaBar Collaboration),
\href{https://journals.aps.org/prd/abstract/10.1103/PhysRevD.76.052005}{Phys. Rev. D \textbf{76}, 052005 (2007).}



\bibitem{Belle:2006idb}
L.~Widhalm \textit{et al.} (Belle Collaboration),
\href{https://journals.aps.org/prl/abstract/10.1103/PhysRevLett.97.061804}{Phys. Rev. Lett. \textbf{97}, 061804 (2006).}



\bibitem{CLEO:2005rxg}
G.~S.~Huang \textit{et al.} (CLEO Collaboration),
\href{https://journals.aps.org/prl/abstract/10.1103/PhysRevLett.95.181801}{Phys. Rev. Lett. \textbf{95}, 181801 (2005).}


\bibitem{CLEO:2005cuk}
T.~E.~Coan \textit{et al.} (CLEO Collaboration),
\href{https://journals.aps.org/prl/abstract/10.1103/PhysRevLett.95.181802}{Phys. Rev. Lett. \textbf{95}, 181802 (2005).}


\bibitem{CLEO:2007ntr}
S.~Dobbs \textit{et al.} (CLEO Collaboration),
\href{https://journals.aps.org/prd/abstract/10.1103/PhysRevD.77.112005}{Phys. Rev. D \textbf{77}, 112005 (2008).}


\bibitem{CLEO:2009svp}
D.~Besson \textit{et al.} (CLEO Collaboration),
\href{https://journals.aps.org/prd/abstract/10.1103/PhysRevD.80.032005}{Phys. Rev. D \textbf{80}, 032005 (2009).}



\bibitem{BESIII:2021mfl}
M.~Ablikim \textit{et al.} (BESIII Collaboration),
\href{https://journals.aps.org/prd/abstract/10.1103/PhysRevD.104.052008}{Phys. Rev. D \textbf{104}, 052008 (2021).}


\bibitem{BESIII:2015tql}
M.~Ablikim \textit{et al.} (BESIII Collaboration),
\href{https://journals.aps.org/prd/abstract/10.1103/PhysRevD.92.072012}{Phys. Rev. D \textbf{92}, 072012 (2015).}


\bibitem{BESIII:2018ccy}
M.~Ablikim \textit{et al.} (BESIII Collaboration),
\href{https://journals.aps.org/prl/abstract/10.1103/PhysRevLett.122.011804}{Phys. Rev. Lett. \textbf{122}, 011804 (2019).}

\bibitem{BESIII:2015jmz}
M.~Ablikim \textit{et al.} (BESIII Collaboration),
\href{https://journals.aps.org/prd/abstract/10.1103/PhysRevD.92.112008}{Phys. Rev. D \textbf{92}, 112008 (2015).}


\bibitem{BESIII:2017ylw}
M.~Ablikim \textit{et al.} (BESIII Collaboration),
\href{https://journals.aps.org/prd/abstract/10.1103/PhysRevD.96.012002}{Phys. Rev. D \textbf{96}, 012002 (2017).}


\bibitem{BESIII:2016hko}
M.~Ablikim \textit{et al.} (BESIII Collaboration),
\href{https://iopscience.iop.org/article/10.1088/1674-1137/40/11/113001}{Chin. Phys. C \textbf{40}, 113001 (2016).}



\bibitem{BESIII:2016gbw}
M.~Ablikim \textit{et al.} (BESIII Collaboration),
\href{https://link.springer.com/article/10.1140/epjc/s10052-016-4198-2}{Eur. Phys. J. C \textbf{76}, 369 (2016).}


\bibitem{Luminosity}
M.~Ablikim \textit{et al.} (BESIII Collaboration),
\href{https://arxiv.org/abs/2406.05827}{arXiv:2406.05827.}


\bibitem{Ablikim:2009aa}
M.~Ablikim {\it et al.} (BESIII Collaboration),
\href{https://www.sciencedirect.com/science/article/pii/S0168900209023870?via\%3Dihub}
{ Nucl.\ Instrum.\ Meth.\ A {\bf 614}, 345 (2010).}

\bibitem{Yu:IPAC2016-TUYA01}
C.~H.~Yu {\it et al.},
\href{http://accelconf.web.cern.ch/ipac2016/papers/tuya01.pdf}
{Proceedings of IPAC2016, Busan, Korea, 2016,
doi:10.18429/JACoW-IPAC2016-TUYA01.}


\bibitem{Ablikim:2019hff}
M.~Ablikim {\it et al.} (BESIII Collaboration),
\href{https://doi.org/10.1088/1674-1137/44/4/040001}
{Chin. Phys. C {\bf 44}, 040001 (2020).}


\bibitem{Li:2021iwf}
H.~B.~Li and X.~R.~Lyu,
\href{https://academic.oup.com/nsr/article/8/11/nwab181/6381732?login=false}
{Natl. Sci. Rev. {\bf 8}, no.11, nwab181 (2021).}


\bibitem{etof}
 X.~Li {\it et al.},
 \href{https://doi.org/10.1007/s41605-017-0014-2}
 {Radiat. Detect. Technol. Methods {\bf 1}, 13 (2017);}
 Y.~X.~Guo {\it et al.},
 \href{https://doi.org/10.1007/s41605-017-0012-4}
 {Radiat. Detect. Technol. Methods {\bf 1}, 15 (2017);}
 P.~Cao {\it et al.},
 \href{https://doi.org/10.1016/j.nima.2019.163053}
 {Nucl.\ Instrum.\ Meth.\ A {\bf 953}, 163053 (2020).}


\bibitem{geant4}
  S.~Agostinelli {\it et al.} (GEANT4 Collaboration),
  \href{https://doi.org/10.1016/S0168-9002(03)01368-8}
  {Nucl.\ Instrum.\ Meth.\ A {\bf 506}, 250 (2003).}

\bibitem{Huang:2022wuo}
K.~X.~Huang, Z.~J.~Li, Z.~Qian, J.~Zhu, H.~Y.~Li, Y.~M.~Zhang, S.~S.~Sun and Z.~Y.~You,
\href{https://link.springer.com/article/10.1007/s41365-022-01133-8}{Nucl. Sci. Tech. {\bf 33}, 142 (2022).}

\bibitem{ref:kkmc}
  S.~Jadach, B.~F.~L.~Ward and Z.~Was,
  \href{https://doi.org/10.1103/PhysRevD.63.113009}
 { Phys.\ Rev.\ D {\bf 63}, 113009 (2001);}
 \href{https://doi.org/10.1016/S0010-4655(00)00048-5}
{  Comput.\ Phys.\ Commun.\  {\bf 130}, 260 (2000).}

\bibitem{Becher:2005bg}
T.~Becher and R.~J.~Hill,
\href{https://linkinghub.elsevier.com/retrieve/pii/S0370269305017235}{Phys.\ Lett.\ B {\bf 633}, 61 (2006).}

 \bibitem{ref:evtgen}
  D.~J.~Lange,
 \href{https://doi.org/10.1016/S0168-9002(01)00089-4}
 { Nucl.\ Instrum.\ Meth.\ A {\bf 462}, 152 (2001);}
  R.~G.~Ping,
\href{https://iopscience.iop.org/article/10.1088/1674-1137/32/8/001}
{  Chin. Phys. C {\bf 32}, 599 (2008).}

\bibitem{ref:lundcharm}
  J.~C.~Chen, G.~S.~Huang, X.~R.~Qi, D.~H.~Zhang and Y.~S.~Zhu,
\href{https://doi.org/10.1103/PhysRevD.62.034003}
 { Phys.\ Rev.\ D {\bf 62}, 034003 (2000);}
  R.~L.~Yang, R.~G.~Ping and H.~Chen,
\href{https://doi.org/10.1088/0256-307X/31/6/061301}
 { Chin.\ Phys.\ Lett.\  {\bf 31}, 061301 (2014).}

 \bibitem{photos}
  E.~Richter-Was,
\href{https://doi.org/10.1016/0370-2693(93)90062-M}
{  Phys.\ Lett.\ B {\bf 303}, 163 (1993).}

\bibitem{DTmethod}
R. M. Baltrusaitis {\it et al.} (MARK III Collaboration),
\href{https://journals.aps.org/prl/abstract/10.1103/PhysRevLett.56.2140}{Phys. Rev. Lett. {\bf 56}, 2140 (1986).}


\bibitem{deltakpi}
M. Ablikim {\it et al.} (BESIII Collaboration),
\href{https://doi.org/10.1016/j.physletb.2014.05.071}{Phys. Lett. B {\bf 734}, 227 (2014).}

\bibitem{RootClass}
K.~Cranmer \textit{et al.} (ROOT Collaboration),
\href{https://inspirehep.net/literature/1236448}{CERN-OPEN-2012-016.}

\bibitem{argus}
H. Albrecht {\it et al.} (ARGUS Collaboration),
\href{https://doi.org/10.1016/0370-2693(90)91293-K}{Phys. Lett. B {\bf 241}, 278 (1990).}


\bibitem{chiV}
C. G. Boyd, B. Grinstein and R. F. Lebed,
\href{https://linkinghub.elsevier.com/retrieve/pii/0550321395006532}{ Nucl. Phys. B {\bf 461}, 493 (1996).}



\bibitem{Riggio:2017zwh}
L.~Riggio, G.~Salerno and S.~Simula,
\href{https://doi.org/10.1140/epjc/s10052-018-5943-5}{Eur. Phys. J. C \textbf{78}, no.6, 501 (2018).}


\end{thebibliography}
\end{document}